\documentclass[usenatbib]{aa}

\usepackage{txfonts}
\usepackage{graphicx}
\usepackage[update,prepend]{epstopdf}
\usepackage{natbib}

\usepackage{color}
\usepackage{amstext}

\begin{document}

\title{BEER analysis of Kepler and CoRoT light curves}

\subtitle{III. Spectroscopic confirmation of seventy new beaming binaries discovered in CoRoT light curves}

\author{
L. Tal-Or \inst{1}
\and S. Faigler\inst{1}
\and T. Mazeh \inst{1}
}

\institute{School of Physics and Astronomy, Raymond and Beverly Sackler Faculty of
Exact Sciences, Tel Aviv University, Tel Aviv, Israel\\
\email{levtalo@post.tau.ac.il}
}

\date{Received ... ; accepted ...}
\abstract
{The BEER algorithm searches stellar light curves for the BEaming, Ellipsoidal, and Reflection photometric modulations that are caused by a short-period companion. These three effects are typically of very low amplitude and can mainly
be detected in light curves from space-based photometers. Unlike eclipsing binaries, these effects are not limited to edge-on inclinations.}
{Applying the algorithm to wide-field photometric surveys such
as CoRoT and \textit{Kepler} offers an opportunity to better understand the statistical properties of short-period binaries. It also widens the window for detecting intrinsically rare systems, such as short-period brown-dwarf and massive-planetary companions to main-sequence stars.
}
{Applying the search to the first five long-run center CoRoT fields, we identified $481$ non-eclipsing candidates with periodic flux amplitudes of $0.5$--$87$\,mmag. Optimizing the Anglo-Australian-Telescope pointing coordinates and the AAOmega fiber-allocations with dedicated softwares, we acquired six spectra for $231$ candidates and seven spectra for another $50$ candidates in a seven-night campaign. Analysis of the red-arm AAOmega spectra, which covered the range of $8342$--$8842$\,\AA{}, yielded a radial-velocity precision of $\sim1$\,km/s. Spectra containing lines of more than one star were analyzed with the two-dimensional correlation algorithm TODCOR.}
{The measured radial velocities confirmed the binarity of seventy of the BEER candidates---$45$ single-line binaries, $18$ double-line binaries, and $7$ diluted binaries. We show that red giants introduce a major source of false candidates and demonstrate a way to improve BEER's performance in extracting higher fidelity samples from future searches of CoRoT light curves.
The periods of the confirmed binaries span a range of $0.3$--$10$\,days and show a rise in the number of binaries per $\Delta$log$P$ toward longer periods. The estimated mass ratios of the double-line binaries and the mass ratios assigned to the single-line binaries, assuming an isotropic inclination distribution, span a range of $0.03$--$1$. On the low-mass end, we have detected two brown-dwarf candidates on a $\sim1$\, day period orbit.}
{This is the first time non-eclipsing beaming binaries are detected in CoRoT data, and we estimate that $\sim300$ such binaries can be detected in the CoRoT long-run light curves.}
\authorrunning{Tal-Or, Faigler, and Mazeh}
\titlerunning{Seventy new CoRoT BEER binaries}
\maketitle

\section{Introduction}
The space-based photometric surveys CoRoT \citep{rouan98,Baglin2003} and \textit{Kepler} \citep{Borucki2010Sci} were designed mainly to detect minute exoplanetary transits. Each of these missions has provided over $160,000$ continuous stellar light curves with time-spans of tens to hundreds of days and a photometric precision of $10^{-3}$--$10^{-4}$ per measurement \citep{auvergne09,Koch2010ApJ}. Hundreds of transiting planets were indeed detected \citep[e.g.,][]{Moutou2013,Rowe2014}, but the unprecedented quality of the light
 curves also enabled the detection of other flux variations of astrophysical origins. One such variation is the relativistic beaming effect.

\citet{RL1979} showed that several factors contribute to the beaming effect, which increase (decrease) the observed brightness of any approaching (receding) light source by $\sim|4v_R/c|$, where $v_R$ is the radial velocity (RV) of the source and $c$ is the speed of light. Thus, for short-period ($1$--$10$\,days) brown-dwarfs (BDs) or massive planetary companions ($M_p\sin i\gtrsim5M_J$) of solar-like stars, the beaming amplitudes are in the range of $10^{-4}$--$10^{-5}$. \citet{Loeb2003} have predicted that \textit{Kepler}'s photometric precision would be sufficient to detect such companions. For stellar binaries with typical RV semi-amplitudes of $10-100$\,km\,s$^{-1}$, the beaming amplitudes are in the range of $10^{-3}$--$10^{-4}$. For this reason, \citet*{zma2007} have predicted that CoRoT and \textit{Kepler} will also detect hundreds of non-eclipsing binaries of this type, and create a new observational category: beaming binaries.

Soon after the first \textit{Kepler} light curves became available, several studies measured the beaming effect of a few eclipsing binaries (EBs) detected by \textit{Kepler} \citep[e.g.,][]{vanKerkwijk2010,Carter2011ApJ}. However, as mentioned by \citet{Loeb2003} and by \citet*{zma2007}, for short-period binaries and planets, the amplitude of the beaming effect might be comparable to (or even smaller than) another two photometric effects---the ellipsoidal and the reflection light variations. The ellipsoidal variation is caused by tidal interactions between the two components of the binary \citep[e.g.,][]{Morris1985,mazeh08}. The reflection variation is caused by the brightness difference between the 'day' side and the 'night' side of each component \citep[e.g.,][]{Wilson1990,Harrison2003}. By accounting for the three effects, several studies succeeded to detect the weak beaming effect caused by a transiting BD or even a transiting massive planet in CoRoT and \textit{Kepler} light curves \citep[e.g.,][]{Mazeh2010,Shporer2011,Mazeh2012,Jackson2012,Mislis2012}.

To find non-eclipsing short-period beaming binaries, \citet{Faigler2011} introduced the BEER algorithm, which searches light curves for a combination of the three photometric effects caused by a short-period companion---the BEaming, Ellipsoidal, and Reflection periodic modulations. BEER approximates each of the three effects by a sine/cosine function relative to phase zero taken at the time of conjunction---when the lighter component is in front of the heavier one. The reflection and the beaming effects can then be approximated by cosine and sine functions with the orbital period, respectively, while the ellipsoidal effect can be approximated by a cosine function with half the orbital period.

Detection of BEER-like modulations in a stellar light curve does not yet prove the binary nature of the star, since sinusoidal flux modulations could be produced by other effects as well \citep[e.g.,][]{aigrain04}. To confirm BEER detections, one needs to perform RV follow-up observations \citep{Faigler2011}. The first RV confirmation of non-eclipsing beaming binaries was reported by \citet{Faigler2012}. Candidate binaries for that study were detected with BEER in Q0-Q2 \textit{Kepler} light curves, and seven of them were confirmed using RV measurements.

Paper I of the current series \citep{Faigler2013} reported the discovery of Kepler-76b---the first hot Jupiter discovered with BEER. In Paper II, \citet{Faigler2015} showed evidence for equatorial superrotation of three hot Jupiters measured by \textit{Kepler}---KOI-13, HAT-P-7, and Kepler-76b.

In this paper we present RV confirmation of seventy new beaming binaries found by BEER in CoRoT light curves. The targets were selected from the first five long-run center CoRoT fields and were confirmed using the AAOmega multi-object spectrograph \citep{Lewis2002a}. Section \ref{sec2} presents the BEER search applied to the CoRoT light curves, Sect. \ref{sec3} describes the spectroscopic follow-up observations, Sect. \ref{sec4} details the spectral analysis and derivation of RVs from the spectra, Sect. \ref{sec5} explains the orbital solutions performed and the statistical methods applied to separate true BEER binaries from false detections, Sect. \ref{sec6} discusses the use of the findings to evaluate the BEER algorithm performance, Sect. \ref{sec7} focuses on the mass ratio and orbital-period distributions of the new binaries, and Sect. \ref{sec8} summarizes the findings.

\section{BEER photometric search for binaries}
\label{sec2}

\begin{figure*}
\resizebox{\hsize}{!}
{\includegraphics{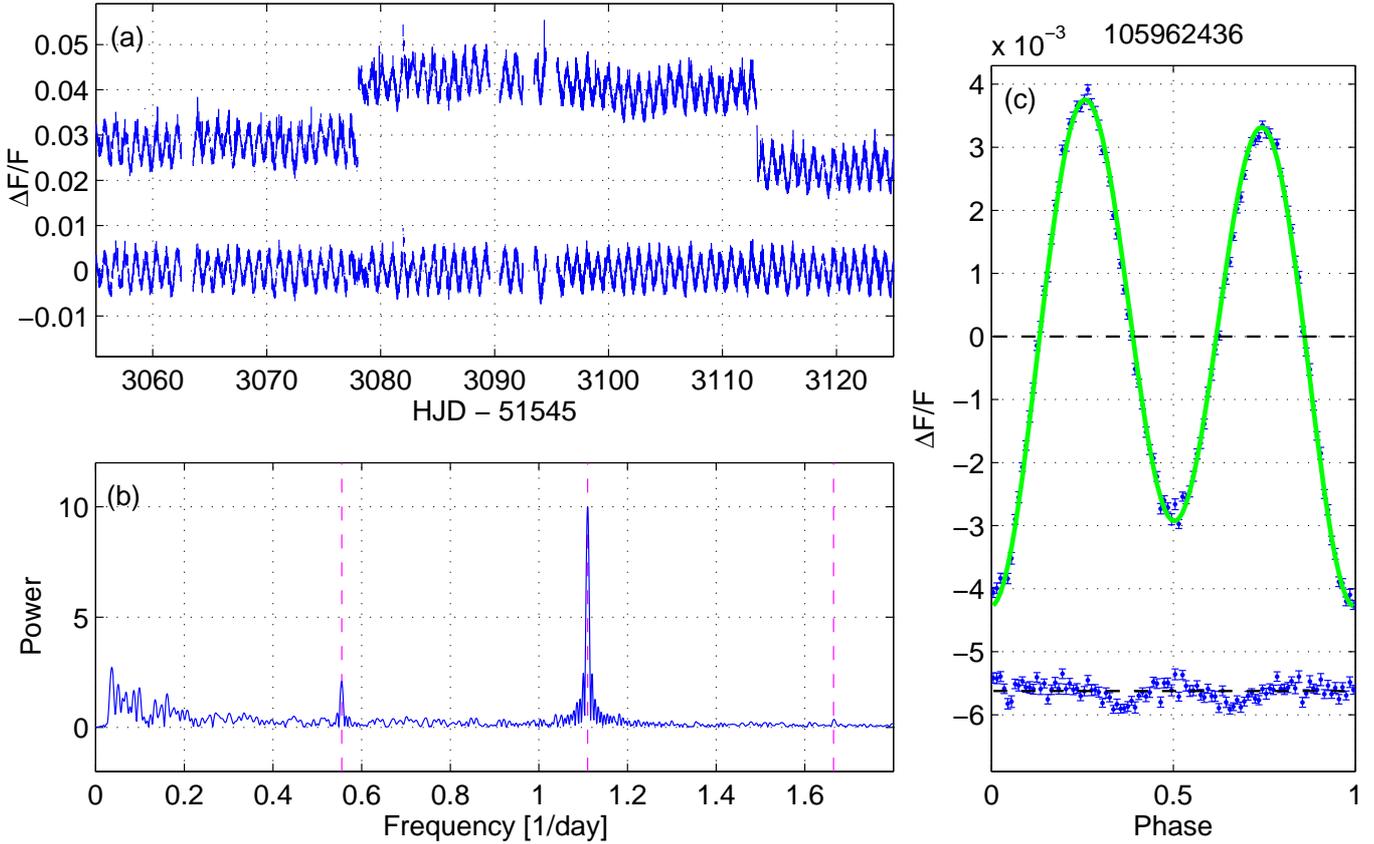}}
\caption{BEER light curve analysis of CoRoT $105962436$ (a) A $70$-day long part of the CoRoT white light curve, normalized by its median. For clarity, the original light curve was shifted upward by $0.03$ relative to the cleaned and detrended one. The cleaned light curve demonstrates the jumps correction, outlier removal, and detrending functionality. (b) FFT-based power spectrum of the detrended light curve, with its maximum value normalized to $10$. 
Vertical dashed lines mark the first three harmonics of the candidate orbital frequency. (c) Phase-folded and binned light curve (blue) and the best-fit BEER models assuming a circular orbit (green). The residuals were shifted downward for clarity.}
\label{Fig1}
\end{figure*}

To detect beaming-binary candidates, we analyzed the $\sim40,000$ white light curves of the CoRoT fields LRc01, LRc02, LRc03, LRc04, and LRc05. We did not use the red, green, or blue light curves of targets having chromatic information \citep[e.g.,][]{Aigrain2008,Deleuil2011} due to their lower signal-to-noise ratio (S/N).

The light curve analysis consisted of several steps. First, oversampled light curves \citep{CoRoTalarm} were rebinned to $512$\,s. Then, we corrected for jumps in all light curves. Jumps were identified by calculating a median filter to each light curve and detecting outliers in its derivative. The correction was made by subtracting the median filter from the light curve around the identified jump epoch. A cosine-transform-based detrending and $4\sigma$ outliers removal were then performed using {\tiny ROBUSTFIT} \citep{HW1997}. Next, a fast Fourier-transform (FFT) -based power spectrum (PS) was calculated for each light curve, and the five most prominent peaks were identified and analyzed following \citet{Faigler2013}. As detailed there, for each of the five peaks we derived the BEER amplitudes and the estimated mass and albedo of the presumed companion by fitting the amplitudes with a BEER model, assuming that the orbit is circular and the peak corresponds to either the orbital period or half the orbital period. Hence, for each light curve we evaluated ten possible orbital periods.

\begin{figure*}
\resizebox{\hsize}{!}
{\includegraphics{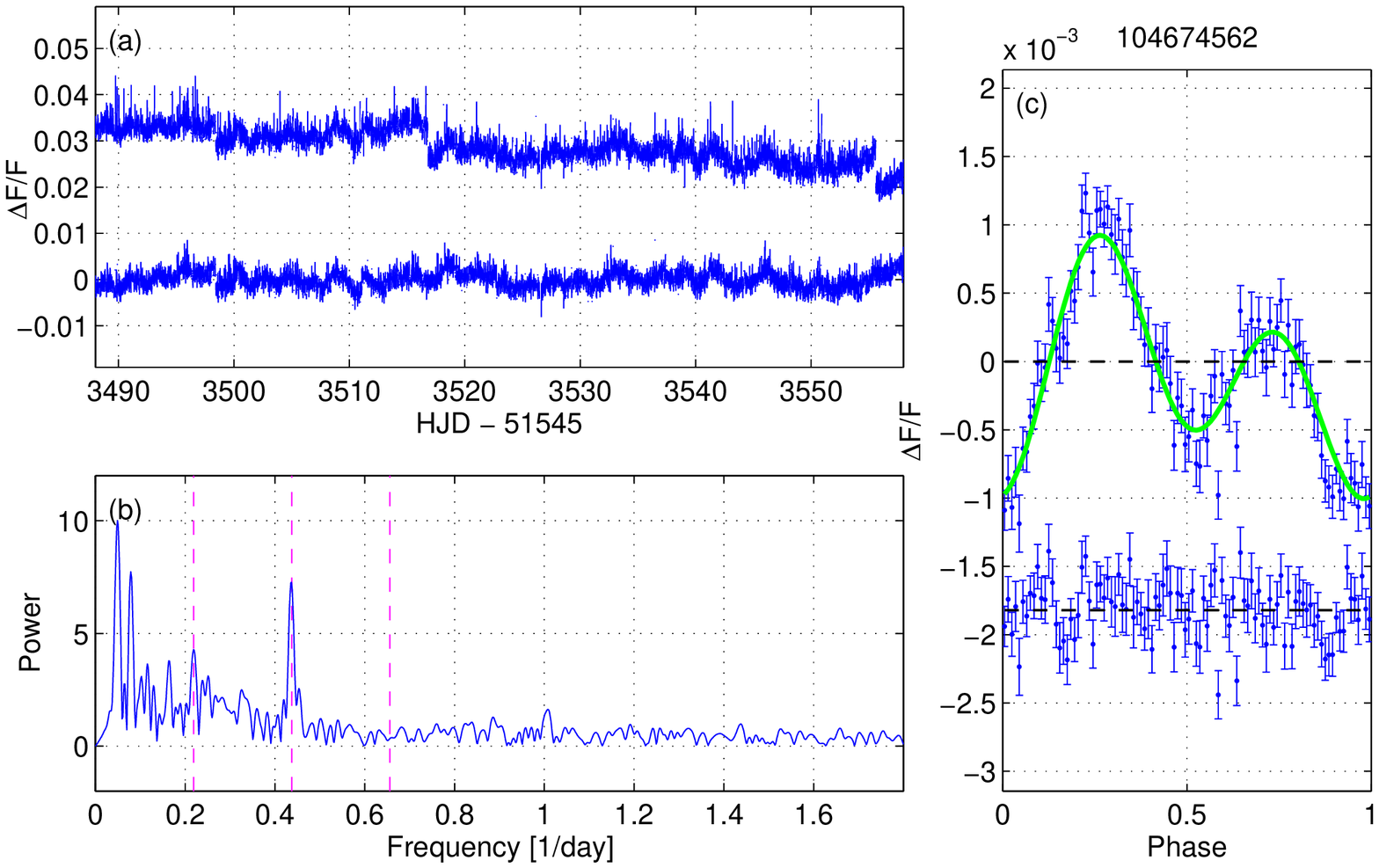}}
\caption{BEER light curve analysis of CoRoT $104674562$. The panels are the same as in Figure \ref{Fig1}.}
\label{Fig2}
\end{figure*}

After fitting a circular BEER model to each light curve at each of its suspected periods, we assigned each fit a score in the $0$--$1$ range, with $1$ being the best. The score of each fit was calculated as
\begin{equation}
S_{\rm Total} = S_{\rm BEER\_S/N} S_{\rm min\_S/N} S_{\chi^2} S_{\rm \sin i} S_{\rm Albedo},
\label{eq:score}
\end{equation}
where $S_{\rm BEER\_S/N}$ is the BEER model S/N score, $S_{\rm min\_S/N}$ is the score of the minimum S/N of the two BEER harmonics, $S_{\chi^2}$ is the fit $\chi^2$ score, $S_{\rm \sin i}$ is the score of the model-derived $\sin i,$ and $S_{\rm albedo}$ is the score of the model-derived geometric albedo.
Each of these partial scores is the result of a dedicated scoring function that gives a score in the range of $0$--$1$ for its associated parameter.
The scoring functions we used were
\begin{equation}
S_{\rm BEER\_S/N} = 1 - \exp \left(-\frac{BEER\_S/N}{C_{\rm BEER\_S/N}}\right),
\label{eq:parscore}
\end{equation}
\begin{equation}
S_{\rm min\_S/N} = 1 - \exp \left(-\frac{min\_S/N}{C_{\rm min\_S/N}}\right),
\end{equation}
\begin{equation}
S_{\chi^2} = \exp \left(-\frac{\chi^2}{C_{\chi^2}}\right),
\end{equation}
\begin{equation}
S_{\rm \sin i} = 1 - \exp \left(-\frac{\sin i}{C_{\rm \sin i}}\right),
 \text{and}\end{equation}
\begin{equation}
S_{\rm Albedo} = \exp \left(-\frac{max(Albedo-C_{\rm alb\_m},0)}{C_{\rm alb\_s}}\right),
\end{equation}
where $C_{\rm BEER\_S/N}$, $C_{\rm min\_S/N}$, $C_{\chi^2}$, $C_{\rm \sin i}$, $C_{\rm alb\_m}$, and $C_{\rm alb\_s}$ are constants that calibrate the behavior of the scoring functions. The most likely orbital period of each light curve was selected as the period with the highest total score $S_{\rm Total}$ out of its ten evaluated periods.

As an illustration, Figs. \ref{Fig1} and \ref{Fig2} show the light curve analysis of CoRoT $105962436$ and CoRoT $104674562$, respectively. We present $70$-day-long parts of the original and detrended light curves, the entire light curve power spectra, and the entire phase-folded and binned light curves, with the best-fit BEER models superposed. We note that the most prominent peak in Fig. \ref{Fig1} at $1.11$\,day$^{-1}$ is caused by the ellipsoidal effect, whose frequency is twice the orbital frequency (i.e., the predicted orbital period is $1.8$\,days). Figure \ref{Fig2} shows that the BEER modulation of CoRoT $104674562$ at a period of $4.6$\,days is almost buried in the noise, and from the PS it seems that the modulation at $\sim20.4$\,days is more evident. Nevertheless, thanks to its period-selection process, BEER found the correct orbital period in this case as well (as we show in Sect. \ref{sec5}).

After the score assignment and the best-period selection, we visually inspected the $200$ highest score candidates of each field and assigned priorities of $1$--$3$ to targets that we deemed viable binary candidates, with priority $1$ assigned to the best candidates. These priorities were assigned through visual inspection of the photometric power spectrum, the goodness-of-fit of the BEER model, the correlation structure of the time-domain residuals, and other target specific features. Naturally, the visual inspection method and the resulting target priorities are subjective and prone to human bias and errors, but we nevertheless used it for lack of a better software-based method at this stage.

In this way, we selected a total of $481$ candidates for RV follow-up from all five fields. The assigned priorities of these candidates were later used by the AAOmega fiber-allocation software to prioritize between them in cases of conflict. The selected candidates span a range of $0.1$--$17$\,days in predicted orbital periods and of $0.3$--$87$\,mmag in photometric amplitudes. As we show below, the confirmed candidates span somewhat narrower ranges of $0.3$--$10$\,days in orbital period and of $0.5$--$87$\,mmag in photometric amplitudes.

\section{AAOmega follow-up observations}
\label{sec3}

We performed RV follow-up observations with the AAOmega multi-object spectrograph \citep[e.g.,][]{Lewis2002a,Smith2004,Saunders2004} at the Anglo-Australian Telescope (AAT). The AAOmega two-degree field of view, its three-magnitudes dynamic range, and its ability to simultaneously record up to $392$ spectra are ideal for our purpose. These features enabled us to observe the majority of BEER candidates in each of the CoRoT fields in a single pointing. We used the AAOmega software {\tiny CONFIGURE} \citep[e.g.,][]{Lewis2002a,Miszalski2006}\footnote{Available at http://www.aao.gov.au/science/software/configure.} to optimize the pointing and fibre allocation. Table \ref{t1} lists for each field the selected pointing ephemeris, the number of BEER targets observed, and the total number of science targets observed. The observed stars span the range of $12.5$--$16$ in V magnitude.

In total, we observed $281$ out of the $481$ selected BEER candidates. Most of the candidates that were selected for follow-up but were not observed are from the LRc01 and LRc02 fields. These two fields were observed by CoRoT with two $1.4^{\circ}\times1.4^{\circ}$ detectors, as opposed to the LRc03, LRc04, and LRc05 fields that were observed by CoRoT with only one such detector \citep{Moutou2013}. Nevertheless, to maximize the number of priority $1$ candidates observed each night, we used only one pointing per field. Table \ref{t1a} lists the coordinates, magnitudes, photometric ephemeris, and amplitudes of the three BEER effects for the candidates observed with AAOmega. For convenience, the confirmed binaries are indicated in the rightmost column of Table \ref{t1a}. To make the best use of available observing resources, we also observed several hundred CoRoT EBs. In this paper, however, we report only on observing non-eclipsing BEER candidates and on confirming seventy of them. We leave the EBs spectra analysis to a separate paper.

\begin{table}
\caption{Selected pointing ephemeris of the five CoRoT fields, and the number of targets observed in each field.}
\begin{tabular}{ccccc}
\hline
\hline
Field & Mean RA & Mean Dec & Science & BEER \\
      & (J2000) & (J2000)  & targets & targets \\
\hline
LRc01 & 19:26:25.34 & +01:12:00.5 & $211$ & $41$ \\
LRc02 & 18:41:52.44 & +06:37:03.9 & $292$ & $55$ \\
LRc03 & 18:32:27.45 & -06:21:52.7 & $362$ & $50$ \\
LRc04 & 18:33:50.89 & +08:49:58.9 & $359$ & $68$ \\
LRc05 & 18:39:19.29 & +04:28:21.6 & $165$ & $67$ \\
\hline
\end{tabular}
\label{t1}
\end{table}

\begin{table*}
\caption{Coordinates and photometric parameters of the BEER candidates observed at AAOmega$^a$}
\begin{tabular}{lcccrrrrrc}
\hline
\hline
CoRoT ID & RA & Dec & V & \multicolumn{1}{c}{Orbital} & \multicolumn{1}{c}{Orbital} & \multicolumn{1}{c}{Ellipsoidal} & \multicolumn{1}{c}{Beaming} & \multicolumn{1}{c}{Reflection} & conf.\\
 & (deg) & (deg) & (mag) & \multicolumn{1}{c}{period} & \multicolumn{1}{c}{phase} & \multicolumn{1}{c}{amplitude} & \multicolumn{1}{c}{amplitude} & \multicolumn{1}{c}{amplitude} & flag$^b$\\
 & & & & \multicolumn{1}{c}{(day)} & \multicolumn{1}{c}{(HJD-2451545)} & \multicolumn{1}{c}{(ppm)} & \multicolumn{1}{c}{(ppm)} & \multicolumn{1}{c}{(ppm)} & \\
\hline
$100537909$ & $290.7709$ & $ 1.2476$ & $13.1$ & $     5.75      $ & $   2828.9    $ & $   -641    $ & $    406    $ & $   -216    $ & $1$ \\
            &            &           &        & $     0.19      $ & $      2.5    $ & $    267    $ & $    207    $ & $    204    $ & \\
$100576007$ & $290.8286$ & $ 1.6783$ & $15.1$ & $ 0.319485      $ & $ 2833.575    $ & $   -902    $ & $    227    $ & $   -187    $ & $0$ \\
            &            &           &        & $ 0.000074      $ & $    0.017    $ & $     27    $ & $     21    $ & $     44    $ & \\
$100604403$ & $290.8710$ & $ 1.6941$ & $13.7$ & $    3.094      $ & $  2832.39    $ & $   -164    $ & $    327    $ & $    120    $ & $0$ \\
            &            &           &        & $    0.021      $ & $     0.50    $ & $     24    $ & $     17    $ & $     26    $ & \\
$100637229$ & $290.9165$ & $ 1.4645$ & $13.8$ & $     8.73      $ & $   2829.7    $ & $   -675    $ & $    480    $ & $   -148    $ & $0$ \\
            &            &           &        & $     0.17      $ & $      1.4    $ & $     45    $ & $     90    $ & $     88    $ & \\
$100677124$ & $290.9786$ & $ 0.9625$ & $15.7$ & $  0.41720      $ & $ 2833.533    $ & $  -9014    $ & $    650    $ & $  -1291    $ & $0$ \\
            &            &           &        & $  0.00015      $ & $    0.026    $ & $     67    $ & $    581    $ & $    420    $ & \\
\hline
\end{tabular}
\\$^{(a)}$\,This table is available in its entirety in a machine-readable form in ftp://wise-ftp.tau.ac.il/pub/corotAAO. A portion of the table is shown here for guidance regarding its form and content. Each line of parameters is followed by a line of uncertainties.
\\$^{(b)}$\,conf. flag: $0$---false alarm; $1$---confirmed BEER binary.
\label{t1a}
\end{table*}

The observations took place on seven consecutive nights, starting on August 02, 2012. Between two and five fields were observed each night, depending on the available time and weather conditions. We acquired six spectra for $231$ candidates and seven spectra for another $50$ candidates in a seven-night AAOmega campaign. Table \ref{t2} lists for each field the heliocentric Julian dates (HJDs) of mid-exposure, calculated for the mean ephemeris presented in Table \ref{t1}.

We used the 1700D grating on the red arm since it was reported to give good RV precision \citep[e.g.,][]{Lane2011}. We used the 1700B grating on the blue arm since it covers several Balmer lines and also enables RV measurements of hot stars. The nominal spectral coverage is $8342$--$8842$\,\AA{} on the red arm and $3803$--$4489$\,\AA{} on the blue arm, but the actual coverage is smaller by up to $\sim60$\,\AA{} and is different for different fibers, depending on their position on the detector.

\begin{table}
\caption{Mid-exposure HJD$-2456141$\,days.}
\begin{tabular}{ccccc}
\hline
\hline
LRc01 & LRc02 & LRc03 & LRc04 & LRc05\\
\hline
$ 1.070956 $ & $    ---   $ & $ 0.998684 $ & $    ---   $ & $    ---   $\\
$ 2.050900 $ & $ 1.942608 $ & $ 1.888283 $ & $ 1.996221 $ & $ 2.103633 $\\
$ 3.045416 $ & $ 2.936347 $ & $ 2.882764 $ & $ 2.989653 $ & $ 3.098236 $\\
$ 4.053797 $ & $ 3.942036 $ & $ 3.887762 $ & $ 3.995172 $ & $ 4.110621 $\\
$ 5.104335 $ & $ 4.942736 $ & $ 4.889214 $ & $ 4.996094 $ & $ 5.049920 $\\
$ 6.109585 $ & $ 5.946059 $ & $ 5.889688 $ & $ 6.001038 $ & $ 6.055340 $\\
$    ---   $ & $ 6.950716 $ & $ 6.899660 $ & $ 7.009869 $ & $ 7.059933 $\\
\hline
\end{tabular}
\label{t2}
\end{table}

Observations and data reduction were performed similarly to previously reported works \citep[see e.g.,][]{Sebastian2012}. In short, the observing sequence at each new pointing usually consisted of a flat and two arc frames followed by $\text{three}$ or $\text{two}$ exposures of $20$ or $30$ minutes, respectively.

Data reduction used the dedicated software {\tiny 2DFDR} \citep{2DFDR}\footnote{Available at http://www.aao.gov.au/science/software/2dfdr.}. The spectrum from each fiber and each subexposure was first normalized by its flat and was wavelength calibrated using the arc frames, rebinning the data onto the same linear wavelength scale to facilitate combining the subexposures. The throughput of each fiber in each subexposure was calculated using sky emission lines. An average sky spectrum was subtracted from the data using dedicated sky fibres. Finally, the subexposures were combined to give a single calibrated $60$-minute exposure, weighting each subexposure by its flux level and rejecting cosmic ray hits.

Most of the exposures were made under bright sky, so the sky subtraction procedure left some residuals of telluric emission lines in the spectra. In addition, a few pixels in each spectrum were affected by bad columns of the detector. We replaced the values of the telluric-line residuals and the bad pixels with a smoothed version of the same spectrum, which was calculated using a moving median filter, $21$-pixel wide. Additional $6\sigma$ outliers were later replaced the same way.

\section{Spectral analysis}
\label{sec4}

\subsection{Non-composite spectra analysis}

\begin{figure}
\resizebox{\hsize}{!}
{\includegraphics{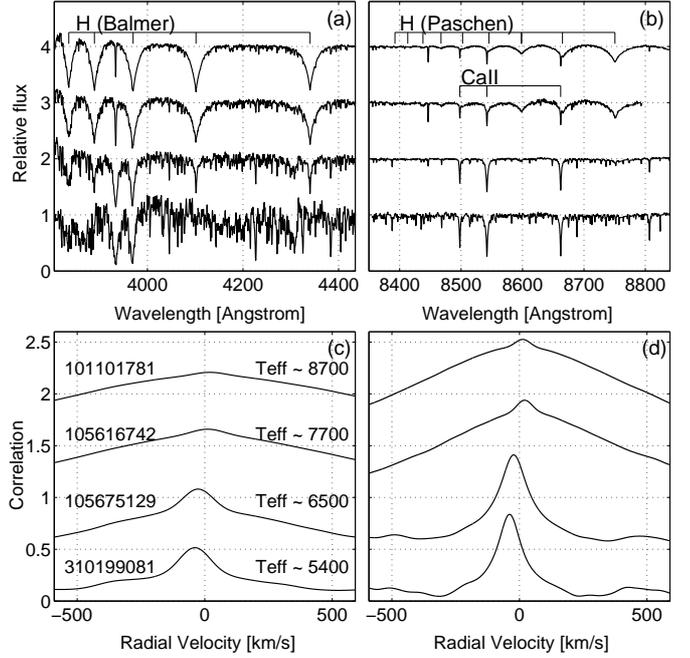}}
\caption{Co-added AAOmega spectra (panels (a) and (b)) and CCFs (panels (c) and (d)) of four BEER candidates with different effective temperatures. For clarity, successive spectra were shifted upward by $1$ and successive CCFs were shifted upward by $0.55$. CoRoT IDs and estimated temperatures of the four candidates are indicated in panel (c). Blue-arm spectra and their CCFs are shown on the left (panels (a) and (c)), while red-arm spectra and their CCFs are shown on the right (panels (b) and (d)). Wavelengths of Ca\,II and hydrogen lines are taken from \citet{NIST_ASD}.}
\label{Fig2_1}
\end{figure}

\begin{figure}
\resizebox{\hsize}{!}
{\includegraphics{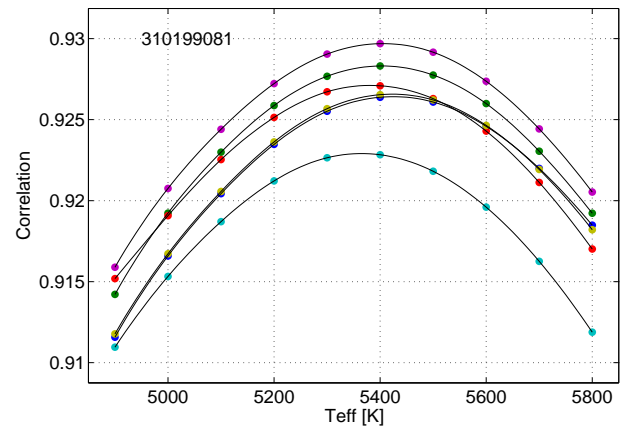}}
\caption{$T_{\rm eff}$ optimization plot of CoRoT $310199081$. Colored dots show the maximum-CCF values, cross-correlating its red-arm AAOmega spectra with synthetic Phoenix spectra of different $T_{\rm eff}$. Solid black lines show for each exposure a {\tiny SPLINE} interpolation of its maximum-CCF versus $T_{\rm eff}$ curve.}
\label{Fig2_2}
\end{figure}

To derive the most precise RVs, we first searched for an optimal theoretical template spectrum for each candidate by maximizing the cross-correlation values between the candidate's observed spectra and a set of synthetic Phoenix spectra \citep{phoenix99}, calculated on a grid of effective temperature ($T_{\rm eff}$), surface gravity (log\,$g$), and metallicity ($[\rm{m/H}]$). Line broadening due to the projected rotational velocity was added to each synthetic spectrum by convolving it with a rotational profile $G(\nu)$ \citep[e.g.,][]{Santerne2012KOI13,Gray2005}. To account for the instrumental broadening of the lines, each synthetic spectrum was also convolved with a Gaussian of $\sigma=20.5$\,km\,s$^{-1}$.

Figures \ref{Fig2_1} and \ref{Fig2_2} illustrate the data and the optimization process. Figure \ref{Fig2_1} shows co-added AAOmega spectra and cross-correlation functions (CCFs) of four slowly rotating candidates (rotational broadening $\lesssim20$\,km\,s$^{-1}$), which have different $T_{\rm eff}$ values and non-composite spectra. The narrow metal lines weaken and wide hydrogen lines become stronger with increasing $T_{\rm eff}$ . Accordingly, the CCF peaks of cool stars ($T_{\rm eff}\lesssim7000$\,K) are narrower than those of hot stars. Figure \ref{Fig2_2} shows the $T_{\rm eff}$ optimization plot of the coolest candidate presented in Fig. \ref{Fig2_1}. The best $T_{\rm eff}$ for each exposure was estimated from the interpolation of its maximum-CCF versus $T_{\rm eff}$ curve. The best global $T_{\rm eff}$ for each candidate was taken as the weighted mean of the different $T_{\rm eff}$ values from the different exposures, taking the squares of the maximum-CCF values as weights. Surface gravity, metallicity, and projected rotational velocity were estimated for each candidate the same way. For each candidate, the nearest synthetic spectrum to its estimated spectral parameters was chosen as its best template.

The analysis, including RV derivation and orbital solutions, was independently performed both for the red-arm and blue-arm spectra. However, the results presented from this point onward are based on the red-arm spectra alone, since we generally found them to have better S/N and stability than the blue-arm spectra. For instance, Fig. \ref{Fig2_1} shows that the blue-arm CCF peaks are wider and lower than the red-arm CCF peaks of the same stars. This is partially due to the lower resolution and lower S/N of blue-arm spectra. Red-arm spectra thus provided better RV precision than blue-arm spectra, even for the hottest stars in the sample ($T_{\rm eff}\gtrsim8000$\,K).

After optimizing the templates, we derived RVs and errors from each spectrum by calculating the CCF with the best template. At this stage, we carefully inspected the CCFs, looking for a signature of a secondary star, and identified $26$ stars that present composite spectra (i.e., spectra containing lines of more than one star). Table \ref{t3a} lists the $255$ non-composite spectrum candidates observed and the template parameters used to derive their RVs. For convenience, the rightmost column identifies the confirmed BEER single-line binaries (SB1s), $45$ of which were identified in that list (as explained in Sect. \ref{sec5}). The measured RVs of all non-composite spectrum candidates are given in machine-readable form at ftp://wise-ftp.tau.ac.il/pub/corotAAO.

\begin{table}
\caption{Template parameters of non-composite spectrum candidates$^a$.}
\begin{tabular}{lccccc}
\hline
\hline
CoRoT & $T_{\rm eff}$ & log\,$g$ & $[\rm{m/H}]$ & Rot. broad. & conf.\\
ID & (K) & (cgs) & (dex) & (km\,s$^{-1}$) & flag\\
\hline
$100537909$ & $6500$ & $5.0$ & $0.0$ & $  7$ & $1$ \\
$100576007$ & $5400$ & $4.0$ & $-0.5$ & $  1$ & $0$ \\
$100604403$ & $7000$ & $4.5$ & $0.0$ & $  9$ & $0$ \\
$100637229$ & $6500$ & $4.5$ & $0.0$ & $ 16$ & $0$ \\
$100677124$ & $6000$ & $4.0$ & $+0.5$ & $102$ & $0$ \\
\hline
\end{tabular}
\\$^{(a)}$\,This table is available in its entirety in machine-readable form at ftp://wise-ftp.tau.ac.il/pub/corotAAO. A portion of the table is shown here for guidance regarding its form and content.
\label{t3a}
\end{table}
 
\subsection{Composite spectra analysis}
 
\begin{figure}
\resizebox{\hsize}{!}
{\includegraphics{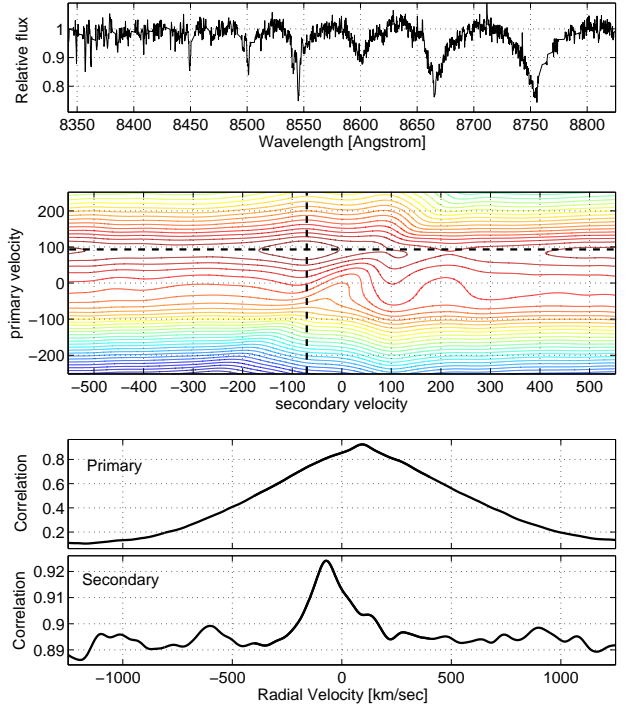}}
\caption{Upper panel: red-arm AAOmega spectrum of the double-line BEER binary $105962436$ from August 05, 2012. Middle: {\tiny TODCOR} two-dimensional correlation function for that spectrum, using the templates whose parameters are specified in Table \ref{t3b}. Colored solid lines connect points of equal correlation. The black dashed lines run through the two-dimensional correlation peak parallel to the primary and secondary RV axes. Bottom panel: primary and secondary cuts through the two-dimensional correlation peak.}
\label{Fig3_2}
\end{figure}

Composite spectra were analyzed with {\tiny TODCOR} \citep{zm94,zmapss94}---the two-dimensional correlation algorithm. The templates were optimized in several steps. First, we searched for the best primary template (i.e., the template for the more luminous component in the spectrum) similarly to the search performed for non-composite spectra. By inspecting the CCFs and the derived RVs, we then identified for each candidate the exposures made at times when the separation between the primary and secondary sets of lines was relatively large. Using only these exposure subsets, we optimized the primary- and secondary-template $T_{\rm eff}$ and rotational broadening, also optimizing the flux ratio between the primary and the secondary components ({\tiny TODCOR}'s $\alpha$ parameter). Metallicity and log\,$g$ of the secondary template (i.e., the template for the less luminous component in the spectrum) for each candidate were fixed assuming the secondary is a dwarf (log\,$g=4.5$) with the same metallicity as its primary.

After optimizing the templates, we used {\tiny TODCOR} to derive the primary and secondary RVs and errors from each spectrum, fixing $\alpha$ to its best value. As an illustration, Fig. \ref{Fig3_2} shows a spectrum and the {\tiny TODCOR} plots for one of the AAOmega exposures of CoRoT $105962436$---probably an A$5$V--G$0$V double-line binary (SB2) with $\alpha\sim0.25$. The splitting of the Ca\,II lines can be seen in the upper panel. The middle panel shows the corresponding {\tiny TODCOR} two-dimensional correlation function. The lower panel of the figure shows the primary and secondary cuts through the two-dimensional correlation function, which run through the two-dimensional correlation peak parallel to the primary and secondary RV axes \citep{zm94}. The correlation in the secondary cut drops only by $\sim0.03$ when moving away from the peak because we changed the velocity of the secondary template, which only contributes $\sim25$\% of the light, and the primary velocity was kept at its best value. Nevertheless, the secondary peak is prominent, which means that the RV of the secondary star can be measured despite the broad hydrogen lines of the primary and the relatively low flux ratio.

\begin{table}
\caption{Template parameters of the composite-spectrum candidates$^a$.}
\begin{tabular}{lccccc}
\hline
\hline
CoRoT & $T_{\rm eff}$ & log\,$g$ & $[\rm{m/H}]$ & Rot. broad. & Flux\\
ID & (K) & (cgs) & (dex) & (km\,s$^{-1}$) & ratio ($\alpha$) \\
\hline
$100688131$\,A & $6100$ & $4.5$ & $0.0$ & $ 13$ & --- \\
$100688131$\,B & $5900$ & $4.5$ & $0.0$ & $  6$ & $0.62$ \\
$100906796$\,A & $7000$ & $4.5$ & $0.0$ & $ 60$ & --- \\
$100906796$\,B & $5100$ & $4.5$ & $0.0$ & $ 71$ & $0.15$ \\
$100976101$\,A & $7400$ & $4.5$ & $0.0$ & $  1$ & --- \\
$100976101$\,B & $4800$ & $4.5$ & $0.0$ & $ 16$ & $0.09$ \\
$101177998$\,A & $7200$ & $4.5$ & $0.0$ & $ 41$ & --- \\
$101177998$\,B & $5000$ & $4.5$ & $0.0$ & $ 14$ & $0.17$ \\
$103833966$\,A & $6200$ & $5.0$ & $+0.5$ & $  1$ & --- \\
$103833966$\,B & $5600$ & $4.5$ & $+0.5$ & $ 21$ & $0.27$ \\
\hline
\end{tabular}
\\$^{(a)}$\,This table is available in its entirety in machine-readable form at ftp://wise-ftp.tau.ac.il/pub/corotAAO. A portion of the table is shown here for guidance regarding its form and content.
\label{t3b}
\end{table}

Table \ref{t3b} lists the template parameters and $\alpha$ values used for the composite-spectrum BEER candidates. We only use the nomenclature 'A' and 'B' (both here and later in Table \ref{t3h}) to denote the more and less luminous components, respectively. The measured RVs and errors of both components are given in machine-readable form at ftp://wise-ftp.tau.ac.il/pub/corotAAO.

\section{Orbital solutions}
\label{sec5}

\subsection{Confirming BEER binaries with non-composite spectra}

To separate true BEER binaries from false detections (which we call false alarms for simplicity, or FAs), and to derive the orbital parameters of the true BEER SB1s, we fitted the derived RVs of non-composite spectrum candidates with a circular Keplerian model. We calculated two $\chi^2$ statistics---$\chi^2_{\rm{null}}$ and $\chi^2_{\rm{orb}}$, for the null hypothesis (constant RV star) and for the circular orbital solution, respectively. We took the BEER period and phase as priors by treating them as additional measurements, meaning that their squares of residuals, scaled by their error estimates, were added to $\chi^2_{\rm{orb}}$ in the search for the best fit. For the best-fit orbital parameters we also calculated the following \textit{F}-statistic:

\begin{equation}
F = \frac{(\chi^2_{\rm{null}}-\chi^2_{\rm{orb}})}{\rm{DOF_{null}}-\rm{DOF_{orb}}}/\frac{\chi^2_{\rm{orb}}}{\rm{DOF_{orb}}} $ , $ 
\label{F}
\end{equation}
where $\rm{DOF_{null}}$ and $\rm{DOF_{orb}}$ are the numbers of degrees of freedom of the null hypothesis and the orbital solution, respectively. Since the only free parameter of the null hypothesis is an RV offset, $\rm{DOF_{null}}=\rm{N_{RV}}-1$, where $\rm{N_{RV}}$ is the number of RV points. For the orbital solution $\rm{DOF_{orb}}=\rm{N_{RV}}+2-p_{\rm{orb}}$, where ${\rm{p_{orb}}}$ is the number of free parameters of the orbital solution, since we treated the BEER period and phase as additional measurements. For a circular orbit ${\rm{p_{orb}}}=4$.

To obtain a good distinction between true BEER SB1s and FAs, we used both the $\chi^2_{\rm{null}}$-test and the \textit{F}-test. The $\chi^2_{\rm{null}}$-test was used first to screen all candidates that showed no significant RV variability within the observing run. Then the \textit{F}-test was applied to candidates that passed the $\chi^2_{\rm{null}}$-test to check the compatibility of their RVs with a circular Keplerian model at the BEER-predicted period and phase. Only candidates that passed both tests were considered as confirmed BEER SB1s.

After visual inspection of the results, particularly of a few borderline cases, we chose the critical \textit{p}-values to be $10^{-6}$ and $0.003$ for the $\chi^2_{\rm{null}}$- and \textit{F}-tests, respectively. As a consequence, $54$ out of the $255$ non-composite spectrum candidates passed the $\chi^2_{\rm{null}}$-test, and $45$ of them also passed the \textit{F}-test and were classified as confirmed BEER SB1s. The $\text{nine}$ candidates that passed the $\chi^2_{\rm{null}}$-test but not the \textit{F}-test might be true variables at a different orbital period, or their spectra suffer from some systematics causing RV outliers.

Another possible reason for a true BEER binary to fail our \textit{F}-test is an eccentric orbit. Therefore, we also fitted each RV curve with an eccentric Keplerian model, for which  ${\rm{p_{orb}}}=6$. The fact we have only $6$--$7$ RV points for most of our candidates makes the $\rm{DOF_{orb}}$ of an eccentric solution as small as $2$--$3$. Since  an \textit{F}-test fails at such a low number of DOF, we required a \textit{p}-value improvement of at least a factor of $10$ to prefer the eccentric solution over the circular one. None of the candidates fulfilled this requirement, meaning we could not find significant eccentricity in any of the confirmed SB1s.

\begin{figure}
\resizebox{\hsize}{!}
{\includegraphics{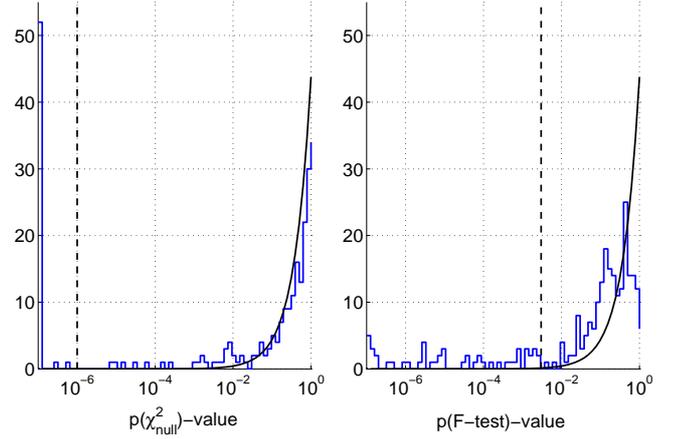}}
\caption{Test statistics \textit{p}-value histograms of the $255$ non-composite spectrum BEER candidates, solving for their red-arm AAOmega RVs. The solid black lines represent a flat distribution between $0$ and $1$ scaled so that its integral is equal the number of FAs ($210$). The dashed black lines mark the critical values that were chosen to separate possible BEER SB1s (to the left of the lines) from FAs (to the right of the lines). The histograms were truncated at $10^{-7}$, so that the leftmost columns contain all candidates with \textit{p}-values $<10^{-7}$.}
\label{Fig3}
\end{figure}

Figure \ref{Fig3} shows (on a log-scale) \textit{p}-value histograms of the $\chi^2_{\rm{null}}$- and \textit{F}-statistics for the $255$ non-composite spectrum BEER candidates. The solid black lines represent the expected \textit{p}-value distributions for RV measurements of constant-RV stars normally distributed for each star around its RV. The expected and observed histograms agree fairly
well, particularly for the $\chi^2_{\rm{null}}$-statistics. The dashed black lines mark the critical values that were chosen to separate possible BEER SB1s from FAs. Only candidates found to the left of the lines in both plots were considered as confirmed BEER SB1s. 

\begin{table*}
\caption{Orbital parameters of the $45$ confirmed BEER SB1s$^a$.}
\begin{tabular}{lrrrrcc}
\hline
\hline
CoRoT ID & \multicolumn{1}{c}{$P$} & \multicolumn{1}{c}{$T_0$} & \multicolumn{1}{c}{$K$} & \multicolumn{1}{c}{$\gamma$} & \textit{F}-test & $\chi^2_{orb}$\\
 & \multicolumn{1}{c}{(day)} & \multicolumn{1}{c}{(HJD-2456141)} & \multicolumn{1}{c}{(km\,s$^{-1}$)} & \multicolumn{1}{c}{(km\,s$^{-1}$)} & \textit{p}-value & \\
\hline
$103924393$ & $    6.851        $ & $  5.906    $ & $49.93      $ & $42.90        $ & $1.3\rm{E}-08$ & $1.6$\\
            & $    0.021        $ & $  0.021    $ & $ 0.75      $ & $ 0.60        $ & & \\
$105767195$ & $   4.0526        $ & $  3.139    $ & $27.99      $ & $ 2.01        $ & $3.4\rm{E}-08$ & $0.7$\\
            & $   0.0022        $ & $  0.019    $ & $ 0.85      $ & $ 0.60        $ & & \\
$105810223$ & $0.5929447        $ & $3.87271    $ & $115.0      $ & $ 4.31        $ & $4.0\rm{E}-08$ & $4.5$\\
            & $0.0000029        $ & $0.00097    $ & $  1.4      $ & $ 0.93        $ & & \\
$104674562$ & $   4.5804        $ & $  5.895    $ & $30.55      $ & $17.73        $ & $4.5\rm{E}-08$ & $0.7$\\
            & $   0.0071        $ & $  0.021    $ & $ 0.96      $ & $ 0.66        $ & & \\
$310222016$ & $  2.08218        $ & $  5.725    $ & $ 29.3      $ & $-30.61       $ & $4.9\rm{E}-08$ & $3.8$\\
            & $  0.00047        $ & $  0.020    $ & $  1.7      $ & $ 0.65        $ & & \\
$310210722$ & $    6.975        $ & $  7.165    $ & $ 47.7      $ & $11.02        $ & $1.1\rm{E}-07$ & $7.6$\\
            & $    0.020        $ & $  0.025    $ & $  1.2      $ & $ 0.78        $ & & \\
$105818861$ & $ 1.301182        $ & $ 6.0599    $ & $ 57.5      $ & $ 9.82        $ & $1.2\rm{E}-07$ & $3.5$\\
            & $ 0.000054        $ & $ 0.0037    $ & $  1.0      $ & $ 0.73        $ & & \\
$101029997$ & $  3.57665        $ & $  3.361    $ & $41.42      $ & $36.17        $ & $1.3\rm{E}-07$ & $2.7$\\
            & $  0.00044        $ & $  0.011    $ & $ 0.88      $ & $ 0.59        $ & & \\
$105844488$ & $ 0.554937        $ & $ 2.9081    $ & $ 90.2      $ & $ -7.7        $ & $1.4\rm{E}-07$ & $2.4$\\
            & $ 0.000095        $ & $ 0.0021    $ & $  2.1      $ & $  1.5        $ & & \\
$310215106$ & $  3.23943        $ & $  4.099    $ & $33.58      $ & $-0.07        $ & $1.7\rm{E}-07$ & $7.4$\\
            & $  0.00032        $ & $  0.016    $ & $ 0.91      $ & $ 0.69        $ & & \\
$310176634$ & $  1.26221        $ & $  5.877    $ & $ 17.1      $ & $-26.82       $ & $3.5\rm{E}-07$ & $0.5$\\
            & $  0.00058        $ & $  0.013    $ & $  1.1      $ & $ 0.81        $ & & \\
$105336757$ & $ 2.356960        $ & $  6.059    $ & $32.78      $ & $-7.64        $ & $3.8\rm{E}-07$ & $3.0$\\
            & $ 0.000096        $ & $  0.011    $ & $ 0.86      $ & $ 0.64        $ & & \\
$105712106$ & $   3.4521        $ & $  7.165    $ & $47.04      $ & $-2.58        $ & $7.7\rm{E}-07$ & $10.4$\\
            & $   0.0020        $ & $  0.010    $ & $ 0.78      $ & $ 0.59        $ & & \\
$105403147$ & $   2.9136        $ & $  5.876    $ & $35.37      $ & $22.14        $ & $1.1\rm{E}-06$ & $7.0$\\
            & $   0.0019        $ & $  0.012    $ & $ 0.79      $ & $ 0.59        $ & & \\
$105706604$ & $  2.26769        $ & $  6.978    $ & $ 32.4      $ & $-17.00       $ & $2.2\rm{E}-06$ & $4.8$\\
            & $  0.00017        $ & $  0.012    $ & $  1.2      $ & $ 0.69        $ & & \\
$105714214$ & $   3.1309        $ & $  1.950    $ & $36.57      $ & $-10.49       $ & $2.4\rm{E}-06$ & $10.0$\\
            & $   0.0051        $ & $  0.012    $ & $ 0.82      $ & $ 0.60        $ & & \\
$104295292$ & $   2.8685        $ & $  2.751    $ & $26.09      $ & $ 7.83        $ & $2.8\rm{E}-06$ & $5.1$\\
            & $   0.0043        $ & $  0.015    $ & $ 0.86      $ & $ 0.60        $ & & \\
$110567660$ & $ 1.238837        $ & $ 3.9782    $ & $ 49.5      $ & $ -8.1        $ & $2.9\rm{E}-06$ & $6.0$\\
            & $ 0.000091        $ & $ 0.0054    $ & $  1.5      $ & $  1.0        $ & & \\
$101177265$ & $0.6419480        $ & $ 4.0489    $ & $ 36.4      $ & $ 35.2        $ & $3.0\rm{E}-06$ & $3.4$\\
            & $0.0000050        $ & $ 0.0038    $ & $  1.5      $ & $  1.0        $ & & \\
$101014035$ & $0.4596816        $ & $ 4.0441    $ & $ 55.3      $ & $  1.9        $ & $4.1\rm{E}-06$ & $7.6$\\
            & $0.0000015        $ & $ 0.0021    $ & $  1.7      $ & $  1.2        $ & & \\
$100889978$ & $   10.160        $ & $  0.734    $ & $27.38      $ & $ 7.84        $ & $5.4\rm{E}-06$ & $9.4$\\
            & $    0.026        $ & $  0.093    $ & $ 0.80      $ & $ 0.98        $ & & \\
$105661774$ & $  2.22043        $ & $  2.094    $ & $ 31.3      $ & $ 26.2        $ & $6.2\rm{E}-06$ & $2.0$\\
            & $  0.00013        $ & $  0.023    $ & $  2.3      $ & $  1.4        $ & & \\
$105618890$ & $   2.6219        $ & $  1.774    $ & $ 39.5      $ & $21.24        $ & $6.5\rm{E}-06$ & $11.0$\\
            & $   0.0012        $ & $  0.013    $ & $  1.1      $ & $ 0.81        $ & & \\
\hline
\end{tabular}
\\$^{(a)}$\,Note: each line of parameters is followed by a line of uncertainties.
\label{t3e}
\end{table*}
 
\addtocounter{table}{-1}

\begin{table*}
\caption{Continued.}
\begin{tabular}{lrrrrcc}
\hline
\hline
CoRoT ID & \multicolumn{1}{c}{$P$} & \multicolumn{1}{c}{$T_0$} & \multicolumn{1}{c}{$K$} & \multicolumn{1}{c}{$\gamma$} & \textit{F}-test & $\chi^2_{orb}$\\
 & \multicolumn{1}{c}{(day)} & \multicolumn{1}{c}{(HJD-2456141)} & \multicolumn{1}{c}{(km\,s$^{-1}$)} & \multicolumn{1}{c}{(km\,s$^{-1}$)} & \textit{p}-value & \\
\hline
$105802223$ & $ 1.414011        $ & $  3.826    $ & $ 18.0      $ & $34.40        $ & $7.6\rm{E}-06$ & $2.0$\\
            & $ 0.000040        $ & $  0.014    $ & $  1.2      $ & $ 0.85        $ & & \\
$100880613$ & $ 0.893171        $ & $ 1.9956    $ & $ 36.0      $ & $ 59.9        $ & $8.0\rm{E}-06$ & $8.3$\\
            & $ 0.000018        $ & $ 0.0069    $ & $  1.3      $ & $  1.1        $ & & \\
$105760939$ & $  1.53138        $ & $  3.847    $ & $ 22.7      $ & $36.56        $ & $8.3\rm{E}-06$ & $2.6$\\
            & $  0.00042        $ & $  0.012    $ & $  1.4      $ & $ 0.89        $ & & \\
$104113878$ & $  2.22939        $ & $  3.085    $ & $15.70      $ & $-25.11       $ & $2.5\rm{E}-05$ & $6.5$\\
            & $  0.00024        $ & $  0.024    $ & $ 0.97      $ & $ 0.59        $ & & \\
$103782315$ & $    5.372        $ & $  6.737    $ & $ 15.9      $ & $-46.68       $ & $3.1\rm{E}-05$ & $2.9$\\
            & $    0.012        $ & $  0.070    $ & $  1.3      $ & $ 0.90        $ & & \\
$310169750$ & $0.6321732        $ & $ 2.8879    $ & $ 40.5      $ & $ -3.8        $ & $3.4\rm{E}-05$ & $8.0$\\
            & $0.0000065        $ & $ 0.0058    $ & $  2.0      $ & $  1.5        $ & & \\
$103922738$ & $    7.901        $ & $   7.92    $ & $  6.2      $ & $12.64        $ & $5.3\rm{E}-05$ & $0.5$\\
            & $    0.037        $ & $   0.19    $ & $  1.4      $ & $ 0.87        $ & & \\
$104648865$ & $ 1.446104        $ & $  5.045    $ & $ 32.6      $ & $-21.2        $ & $6.3\rm{E}-05$ & $0.8$\\
            & $ 0.000097        $ & $  0.041    $ & $  5.9      $ & $  4.3        $ & & \\
$104536524$ & $  3.12477        $ & $  2.108    $ & $ 10.8      $ & $60.34        $ & $6.7\rm{E}-05$ & $2.6$\\
            & $  0.00100        $ & $  0.052    $ & $  1.1      $ & $ 0.79        $ & & \\
$101058035$ & $0.7647344        $ & $ 6.1885    $ & $ 21.3      $ & $10.57        $ & $8.6\rm{E}-05$ & $9.1$\\
            & $0.0000057        $ & $ 0.0079    $ & $  1.3      $ & $ 0.95        $ & & \\
$105378453$ & $   3.2362        $ & $  5.099    $ & $19.77      $ & $-14.99       $ & $1.5\rm{E}-04$ & $23.1$\\
            & $   0.0075        $ & $  0.021    $ & $ 0.83      $ & $ 0.58        $ & & \\
$105597526$ & $ 0.920035        $ & $  2.934    $ & $ 39.0      $ & $-55.7        $ & $2.0\rm{E}-04$ & $7.1$\\
            & $ 0.000029        $ & $  0.023    $ & $  3.7      $ & $  5.8        $ & & \\
$104667709$ & $0.6984611        $ & $ 3.0336    $ & $ 25.1      $ & $-34.8        $ & $3.5\rm{E}-04$ & $13.1$\\
            & $0.0000056        $ & $ 0.0071    $ & $  1.9      $ & $  1.1        $ & & \\
$105154613$ & $    4.587        $ & $  1.459    $ & $ 9.36      $ & $38.69        $ & $5.0\rm{E}-04$ & $8.6$\\
            & $    0.026        $ & $  0.064    $ & $ 0.88      $ & $ 0.59        $ & & \\
$100537909$ & $   4.8190        $ & $  2.534    $ & $24.41      $ & $31.81        $ & $5.6\rm{E}-04$ & $67.6$\\
            & $   0.0094        $ & $  0.023    $ & $ 0.83      $ & $ 0.56        $ & & \\
$105659320$ & $ 0.706141        $ & $  4.986    $ & $ 6.91      $ & $-1.18        $ & $6.7\rm{E}-04$ & $4.7$\\
            & $ 0.000026        $ & $  0.018    $ & $ 0.94      $ & $ 0.72        $ & & \\
$101044188$ & $ 1.368287        $ & $  4.969    $ & $ 8.19      $ & $39.32        $ & $7.1\rm{E}-04$ & $6.9$\\
            & $ 0.000075        $ & $  0.021    $ & $ 0.94      $ & $ 0.62        $ & & \\
$104279119$ & $   10.161        $ & $   7.63    $ & $ 14.8      $ & $ 12.5        $ & $7.3\rm{E}-04$ & $15.2$\\
            & $    0.051        $ & $   0.24    $ & $  1.5      $ & $  1.7        $ & & \\
$104598628$ & $  2.75207        $ & $  4.264    $ & $ 19.1      $ & $ 96.2        $ & $1.3\rm{E}-03$ & $6.5$\\
            & $  0.00061        $ & $  0.056    $ & $  2.8      $ & $  1.8        $ & & \\
$105164611$ & $  3.50481        $ & $  4.532    $ & $ 23.8      $ & $  4.0        $ & $1.4\rm{E}-03$ & $9.0$\\
            & $  0.00066        $ & $  0.056    $ & $  2.9      $ & $  1.8        $ & & \\
$100851348$ & $ 0.854818        $ & $  4.027    $ & $ 13.5      $ & $ 19.0        $ & $1.7\rm{E}-03$ & $4.0$\\
            & $ 0.000012        $ & $  0.023    $ & $  2.5      $ & $  2.0        $ & & \\
$105472536$ & $0.4066543        $ & $ 6.8778    $ & $ 34.0      $ & $-36.8        $ & $2.1\rm{E}-03$ & $4.8$\\
            & $0.0000092        $ & $ 0.0050    $ & $  6.4      $ & $  2.4        $ & & \\
\hline
\end{tabular}
\label{t3e}
\end{table*}

Table \ref{t3e} lists the orbital parameters of the confirmed BEER SB1s. Figure \ref{Fig6} shows their measured AAOmega RVs and the best-fit circular Keplerian model. Their orbital periods span a range of $0.4$--$10$\,days, and their RV semi-amplitudes span a range of $6$--$115$\,km\,s$^{-1}$. Two of the confirmed BEER SB1s with the smallest RV semi-amplitudes are CoRoT $105659320$ and $101044188$---possibly two BDs on a $\sim1$\, -day period orbit around F-G stars.

\subsection{Confirming BEER binaries with composite spectra}

\begin{table*}[!htp]
\caption{Orbital parameters of the $18$ confirmed BEER SB2s$^a$.}
\begin{tabular}{lrrrrrccc}
\hline
\hline
CoRoT ID & \multicolumn{1}{c}{$P$} & \multicolumn{1}{c}{$T_0$} & \multicolumn{1}{c}{$K_1$} & \multicolumn{1}{c}{$K_2$} & \multicolumn{1}{c}{$\gamma$} & \textit{F}-test & \textit{F}-test & $\chi^2_{orb}$\\
 & \multicolumn{1}{c}{(day)} & \multicolumn{1}{c}{(HJD-2456141)} & \multicolumn{1}{c}{(km\,s$^{-1}$)} & \multicolumn{1}{c}{(km\,s$^{-1}$)} & \multicolumn{1}{c}{(km\,s$^{-1}$)} & \textit{p}-value A & \textit{p}-value B & \\
\hline
$105962436$ & $   1.8020        $ & $  5.917    $ & $ 83.5      $ & $133.8        $ & $27.74      $ & $5.6\rm{E}-09$ & $8.0\rm{E}-05$ & $20.3$ \\
            & $   0.0062        $ & $  0.012    $ & $  1.2      $ & $  3.0        $ & $ 0.78      $ & & & \\
$106024478$ & $    3.013        $ & $ 4.1436    $ & $65.25      $ & $ 66.2        $ & $-22.33     $ & $2.0\rm{E}-07$ & $5.4\rm{E}-08$ & $9.9$ \\
            & $    0.012        $ & $ 0.0068    $ & $ 0.98      $ & $  1.1        $ & $ 0.57      $ & & & \\
$105649738$ & $    3.604        $ & $ 2.7189    $ & $ 87.6      $ & $ 87.7        $ & $-7.47      $ & $2.0\rm{E}-07$ & $2.2\rm{E}-07$ & $8.4$ \\
            & $    0.010        $ & $ 0.0064    $ & $  1.1      $ & $  1.2        $ & $ 0.57      $ & & & \\
$105928477$ & $   1.6509        $ & $ 4.9285    $ & $ 74.3      $ & $ 74.7        $ & $-2.18      $ & $2.3\rm{E}-07$ & $1.6\rm{E}-06$ & $3.1$ \\
            & $   0.0048        $ & $ 0.0063    $ & $  1.2      $ & $  2.3        $ & $ 0.80      $ & & & \\
$100688131$ & $    6.984        $ & $  2.518    $ & $38.12      $ & $43.20        $ & $-13.84     $ & $1.7\rm{E}-06$ & $2.3\rm{E}-06$ & $25.1$ \\
            & $    0.098        $ & $  0.023    $ & $ 0.77      $ & $ 0.95        $ & $ 0.47      $ & & & \\
$310198235$ & $    1.713        $ & $  6.873    $ & $ 99.7      $ & $113.4        $ & $ 7.55      $ & $2.0\rm{E}-06$ & $6.3\rm{E}-07$ & $74.8$ \\
            & $    0.011        $ & $  0.019    $ & $  2.8      $ & $  3.7        $ & $ 0.61      $ & & & \\
$100976101$ & $    4.738        $ & $  0.856    $ & $67.22      $ & $113.0        $ & $-9.00      $ & $2.3\rm{E}-06$ & $2.2\rm{E}-05$ & $68.7$ \\
            & $    0.025        $ & $  0.018    $ & $ 0.73      $ & $  2.9        $ & $ 0.55      $ & & & \\
$104369937$ & $    6.093        $ & $  5.699    $ & $56.37      $ & $ 72.8        $ & $-16.58     $ & $2.6\rm{E}-06$ & $2.2\rm{E}-04$ & $54.7$ \\
            & $    0.061        $ & $  0.020    $ & $ 0.81      $ & $  1.7        $ & $ 0.54      $ & & & \\
$105963904$ & $    6.048        $ & $  5.062    $ & $67.26      $ & $ 81.7        $ & $-36.46     $ & $2.8\rm{E}-06$ & $4.7\rm{E}-05$ & $99.9$ \\
            & $    0.044        $ & $  0.010    $ & $ 0.76      $ & $  1.5        $ & $ 0.53      $ & & & \\
$104181232$ & $    4.138        $ & $  4.025    $ & $58.15      $ & $ 83.2        $ & $-0.64      $ & $4.0\rm{E}-06$ & $1.9\rm{E}-04$ & $34.1$ \\
            & $    0.027        $ & $  0.012    $ & $ 0.87      $ & $  3.5        $ & $ 0.65      $ & & & \\
$105506915$ & $   0.6542        $ & $  3.083    $ & $25.50      $ & $123.4        $ & $-72.51     $ & $2.0\rm{E}-05$ & $4.8\rm{E}-05$ & $22.8$ \\
            & $   0.0048        $ & $  0.016    $ & $ 0.98      $ & $  5.0        $ & $ 0.96      $ & & & \\
$310136399$ & $    6.146        $ & $  2.826    $ & $ 73.2      $ & $ 76.0        $ & $-7.10      $ & $3.3\rm{E}-05$ & $2.0\rm{E}-05$ & $96.8$ \\
            & $    0.098        $ & $  0.027    $ & $  1.9      $ & $  1.7        $ & $ 0.97      $ & & & \\
$101177998$ & $   1.6103        $ & $ 4.0305    $ & $ 26.5      $ & $156.8        $ & $40.59      $ & $2.8\rm{E}-04$ & $2.4\rm{E}-04$ & $32.7$ \\
            & $   0.0044        $ & $ 0.0056    $ & $  1.7      $ & $  4.1        $ & $ 0.91      $ & & & \\
$310212616$ & $  0.35918        $ & $ 3.8372    $ & $ 96.7      $ & $149.0        $ & $-12.0      $ & $3.4\rm{E}-04$ & $1.1\rm{E}-05$ & $105.1$ \\
            & $  0.00047        $ & $ 0.0016    $ & $  3.6      $ & $  3.7        $ & $  2.0      $ & & & \\
$103838038$ & $    6.238        $ & $  4.758    $ & $42.98      $ & $ 44.0        $ & $-8.99      $ & $3.8\rm{E}-04$ & $4.4\rm{E}-05$ & $192.2$ \\
            & $    0.061        $ & $  0.016    $ & $ 0.88      $ & $  1.0        $ & $ 0.48      $ & & & \\
$310173237$ & $  0.29882        $ & $ 2.0277    $ & $ 85.5      $ & $132.2        $ & $-38.6      $ & $4.7\rm{E}-04$ & $3.9\rm{E}-04$ & $177.7$ \\
            & $  0.00033        $ & $ 0.0025    $ & $  3.9      $ & $  4.6        $ & $  2.1      $ & & & \\
$104432741$ & $  0.43955        $ & $ 4.0074    $ & $ 86.5      $ & $166.1        $ & $  9.7      $ & $6.0\rm{E}-04$ & $1.3\rm{E}-05$ & $29.8$ \\
            & $  0.00053        $ & $ 0.0018    $ & $  2.7      $ & $  4.1        $ & $  2.0      $ & & & \\
$105583867$ & $   0.8567        $ & $  2.049    $ & $ 54.4      $ & $141.5        $ & $ 10.2      $ & $6.1\rm{E}-04$ & $1.4\rm{E}-04$ & $47.2$ \\
            & $   0.0037        $ & $  0.017    $ & $  2.7      $ & $  7.0        $ & $  2.0      $ & & & \\
\hline
\end{tabular}
\\$^{(a)}$\,Each line of parameters is followed by a line of uncertainties.
\label{t3f}
\end{table*}

There are three possible scenarios
for a composite-spectrum candidate. (1) The two components in the candidate's spectra belong to the primary and secondary stars in a short-period SB2 at the BEER-predicted period and phase. (2) One of the components belongs to a binary at the BEER-predicted period and phase, while the other component belongs to another star, either bound or unbound to the binary (i.e., a diluted BEER binary). (3) Neither of the components belongs to the BEER-predicted binary. We classified a composite-spectrum BEER candidate as confirmed if the RVs of at least one of its components was compatible with a Keplerian model at the BEER-predicted period and phase.

To assign the correct scenario to each composite-spectrum candidate and (in case of a true BEER binary) to derive its orbital parameters, we separately fitted a circular Keplerian model to its primary and secondary RVs and also calculated the \textit{p}-value of the \textit{F}-test in Equation \ref{F}. Candidates in which both component's SB1-model got a \textit{p}-value $<0.001$ were considered as confirmed BEER SB2s. If the SB1 model of only one of the components got a \textit{p}-value $<0.001$, the candidate was considered as a diluted BEER binary. If both component's SB1-model got a \textit{p}-value $>0.001,$ we considered the candidate as an FA. Similarly to non-composite spectrum candidates, the selected critical \textit{p}-value of $0.001$ originates in a visual inspection of the results, particularly of a few borderline cases.

In our sample of $26$ composite-spectrum candidates we found $18$ BEER SB2s, $7$ diluted BEER binaries, and one FA (CoRoT $310186704$). For the confirmed BEER SB2s we then also fitted  circular SB2 Keplerian model for the two sets of RVs together. Together with the $45$ confirmed BEER SB1s, we have thus confirmed $70$ new non-eclipsing BEER binaries.

\begin{table*}[!htp]
\caption{Orbital parameters of the variable component in the
$\text{seven}$ confirmed diluted BEER binaries$^a$.}
\begin{tabular}{lrrrrcc}
\hline
\hline
CoRoT ID & \multicolumn{1}{c}{$P$} & \multicolumn{1}{c}{$T_0^b$} & \multicolumn{1}{c}{$K$} & \multicolumn{1}{c}{$\gamma$} & \textit{F}-test & $\chi^2_{orb}$\\
 & \multicolumn{1}{c}{(day)} & \multicolumn{1}{c}{(HJD-2456141)} & \multicolumn{1}{c}{(km\,s$^{-1}$)} & \multicolumn{1}{c}{(km\,s$^{-1}$)} & \textit{p}-value & \\
\hline
$310205770$\,B & $    5.543     $ & $  5.218    $ & $ 56.8      $ & $ 3.51        $ & $4.9\rm{E}-07$ & $12.6$ \\
             & $    0.014       $ & $  0.019    $ & $  1.5      $ & $ 0.94        $ & & \\
$310193013$\,B & $0.2801751     $ & $ 5.8446    $ & $193.3      $ & $ -0.3        $ & $1.5\rm{E}-05$ & $32.0$ \\
             & $0.0000021       $ & $ 0.0013    $ & $  3.9      $ & $  3.3        $ & & \\
$104791410$\,B & $  1.04457     $ & $  6.757    $ & $  108      $ & $    0        $ & $2.0\rm{E}-05$ & $5.2$ \\
             & $  0.00051       $ & $  0.021    $ & $   18      $ & $   18        $ & & \\
$104626523$\,A & $    7.354     $ & $   2.99    $ & $ 28.2      $ & $  4.7        $ & $1.1\rm{E}-03$ & $0.1$ \\
             & $    0.052       $ & $   0.22    $ & $  1.6      $ & $  1.5        $ & & \\
$105423352$\,B & $    4.712     $ & $  3.668    $ & $ 56.5      $ & $-24.0        $ & $2.6\rm{E}-04$ & $10.8$ \\
             & $    0.018       $ & $  0.040    $ & $  2.5      $ & $  1.8        $ & & \\
$100906796$\,B & $0.8974124     $ & $ 5.0351    $ & $180.0      $ & $ -6.9        $ & $3.5\rm{E}-04$ & $39.7$ \\
             & $0.0000046       $ & $ 0.0043    $ & $  4.8      $ & $  3.2        $ & & \\
$103833966$\,A & $    10.80     $ & $   7.75    $ & $   72      $ & $    1        $ & $4.0\rm{E}-03$ & $0.4$ \\
             & $     0.11       $ & $   0.92    $ & $   47      $ & $   22        $ & & \\
\hline
\end{tabular}
\\$^{(a)}$\,Each line of parameters is followed by a line of uncertainties.
\\$^{(b)}$\,For the eccentric binaries $104626523$\,A and $103833966$\,A the third column shows the periastron time. The eccentricity of $104626523$\,A is $0.33\pm0.09$ and its longitude of periastron is $242\pm15$\,deg. The eccentricity of $103833966$\,A is $0.32\pm0.16$ and its longitude of periastron is $121\pm40$\,deg.
\label{t3h}
\end{table*}

In addition to a circular model, we also fitted each RV curve with an eccentric Keplerian model. Requiring an improvement of a factor $10$  in the \textit{F}-test \textit{p}-value, we found no SB2s that show measurable eccentricity. We found two diluted-binary candidates ($104626523$\,A and $103833966$\,A) to have slightly eccentric orbits. However, the low eccentricities ($\sim0.3$) found for these two cases might also be spurious or at least inflated \citep[e.g.,][]{Lucy1971}. The preference for an eccentric orbital solution might also be a result of systematic RV errors caused by the presence of the second component in the spectra.

Table \ref{t3f} lists the orbital parameters of the confirmed BEER SB2s, sorted by ascending \textit{F}-test \textit{p}-value of the primary. Table \ref{t3h} lists the orbital parameters of the variable components in the diluted BEER binaries, sorted by ascending \textit{F}-test \textit{p}-value.

Figure \ref{Fig7} shows the measured AAOmega RVs and the best-fit Keplerian models of the confirmed BEER SB2s listed in Table \ref{t3f}. Figure \ref{Fig8} shows the measured AAOmega RVs and the best-fit Keplerian models of the variable components in the confirmed BEER diluted binaries listed in Table \ref{t3h}.

Figure \ref{Fig0} shows the phase-folded and binned light curves of all $70$ confirmed BEER binaries, together with the best-fit circular BEER model. For convenience, the order of the plots in Fig. \ref{Fig0} is the same as in Figs. \ref{Fig6}--\ref{Fig8}.

\section{Performance of the BEER search algorithm}
\label{sec6}

We are now in a position to evaluate the performance of the BEER algorithm in detecting short-period binaries in the light curves of CoRoT long runs. This is possible in view of the large sample of confirmations and FAs that are reported here.

\subsection{BEER-model priority, $M_2\sin i$, and period}

We start by considering the priority classes that we manually assigned to candidates during the visual inspection stage. The left-hand side of Table \ref{tab:pri} lists the number of binary confirmations and FAs per priority class within our sample. As expected, BEER did not perform that well with its priority 3 candidates. They were not considered to be good candidates in the first place and were included in the observational campaign only due to the availability of fibers on the AAOmega spectrograph. Therefore, we decided to ignore priority $3$ candidates in our performance analysis, and only concentrated on priority $1$ and $2$ candidates.

\begin{table*}
\caption{Confirmations and FAs per priority class by applying different filters to the sample$^a$.}
\begin{tabular}{c|cc|cc|cc}
\hline
\hline
Filter: & \multicolumn{2}{|c|}{All candidates}  & \multicolumn{2}{|c|}{$M_2\sin i > 0.25 M_{\odot}$} & \multicolumn{2}{|c}{$M_2\sin i > 0.25 M_{\odot}$ and spectral}\\
          & \multicolumn{2}{|c|}{ }               &  \multicolumn{2}{|c|}{ }                           & \multicolumn{2}{|c}{type earlier than G$7$ for $P>1.4$\,day} \\
Priority & Confirmations & FAs & Confirmations & FAs & Confirmations & FAs \\
\hline
$1$ & $30$ & $26$ & $30$ & $18$ & $30$ & $17$ \\
$2$ & $35$ & $108$ & $35$ & $71$ & $34$ & $54$\\
$3$ & $5$ & $77$ & $5$ & $70$ & $4$ & $35$ \\
 \hline
\end{tabular}
\\$^{(a)}$\,See text for the justifications of the chosen filters.
\label{tab:pri}
\end{table*}

\begin{figure}
\resizebox{\hsize}{!}
{\includegraphics{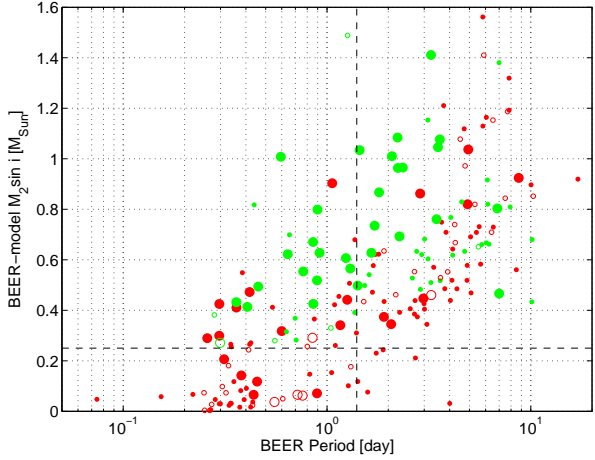}}
\caption{BEER-model $M_2\sin i$ as a function of the BEER period. Large and small circles represent priority $1$ and  $2$ targets, respectively. Green circles represent confirmations, while red circles represent FAs. Open circles represent candidates of spectral type later than G$7$. The horizontal dashed line marks $M_2\sin i=0.25 M_{\odot}$, below which there are no confirmed targets. 
The vertical dashed line marks $P=1.4$\,days, to the right of which there are $18$ FAs of spectral type later than G$7$, but only one confirmed binary (see Sect. \ref{6.2}).}
\label{fig:m2sini}
\end{figure}

Two other parameters that determine the success of the BEER detection are the secondary mass and the orbital period because the BEER amplitudes mostly depend on these two parameters \citep{Faigler2011}. We therefore plot in Fig. \ref{fig:m2sini} the BEER-model $M_2\sin i$ as a function of the photometric period for priority $1$ and $2$ targets. As expected, shorter-period modulations with larger BEER-model $M_2\sin i$ have higher chances of being true BEER modulations than FAs. For instance, there are no confirmations with BEER-model $M_2\sin i < 0.25 M_{\odot}$. The vertical dashed line in Fig. \ref{fig:m2sini} marks $P=1.4$\,days, and it is explained next.

From this stage onward we continue the performance analysis while ignoring all targets with BEER-model $M_2\sin i < 0.25 M_{\odot}$. The summary of the remaining targets as a function of their priority is listed in the central columns of Table \ref{tab:pri}.

\subsection{Spectral type}
\label{6.2}

\begin{figure}
\resizebox{\hsize}{!}
{\includegraphics{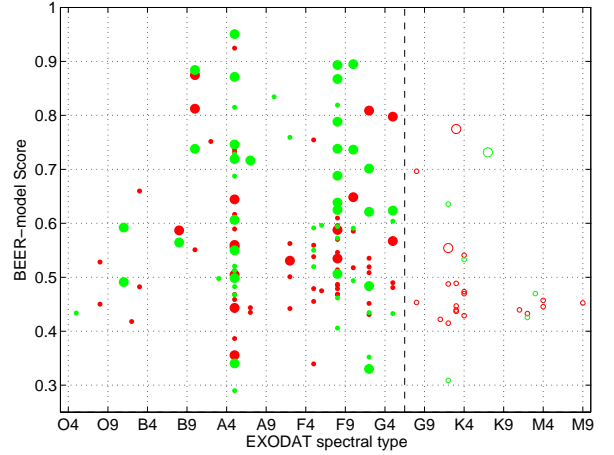}}
\caption{BEER score as a function of the EXODAT spectral type for candidates with $M_2\sin i > 0.25 M_{\odot}$. The symbols are the same as in Fig. \ref{fig:m2sini}. The vertical dashed line marks a spectral type of G$7$.
}
\label{EXODAT}
\end{figure}

Another parameter that can influence the success of the BEER detection is the candidate's spectral type. To check this option, we plot in Fig. \ref{EXODAT} the BEER score as a function of the spectral type for priority $1$ and $2$ candidates with BEER-model $M_2\sin i > 0.25 M_{\odot}$. The spectral type is taken from EXODAT\footnote{cesam.oamp.fr/exodat}, and it was obtained using the SED analysis described by \citet{ExoDatDeleuil2009}. The advantage of using EXODAT is its availability regardless of any follow-up observations. Fig.  \ref{EXODAT} shows that there is a larger fraction of FAs for late-type candidates than for early-type candidates. Particularly if we draw a line at a spectral type of G$7$ ($T_{\rm eff}\sim5600$\,K), we find that the fraction of FAs is $\sim54$\% ($69/128$) to the left of that line and $\sim77$\% ($20/26$) to the right of that line.

In Fig. \ref{fig:m2sini} we also mark with open circles candidates of spectral type later than G$7$. A close examination of these cool candidates, with BEER-model $M_2\sin i > 0.25 M_{\odot}$, reveals that the confirmed binaries differ from the FAs in yet another way. By drawing a line at $P=1.4$\,days, we find most of the cool FAs ($18/20$) to the right of that line, and most of the confirmed cool binaries ($5/6$) to the left of that line.

\begin{figure}
\resizebox{\hsize}{!}
{\includegraphics{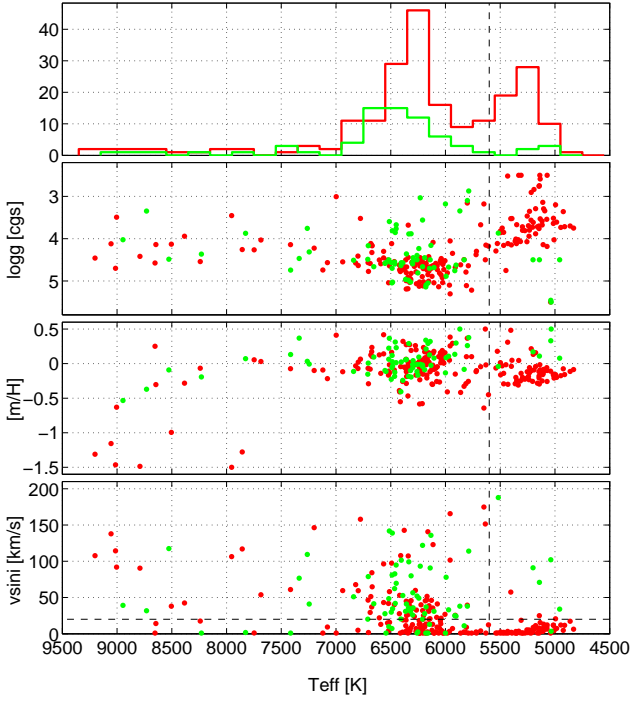}}
\caption{Spectral parameters derived in Sect. \ref{sec4}. Confirmed BEER binaries are represented in green, while FAs are represented in red. For SB2s, only the spectral parameters of the primary are shown. For diluted BEER binaries, only the spectral parameters of the variable component are shown. The upper panel shows the two $T_{\rm eff}$ histograms. The lower three panels show the $T_{\rm eff}$-log\,$g$, $T_{\rm eff}$-metallicity, and $T_{\rm eff}$-rotational broadening scatter plots. The vertical dashed lines mark $T_{\rm eff}=5600$\,K---the approximate temperature of a G$7$V star. The horizontal dashed line in the lower panel indicates a rotational broadening of $20$\,km\,s$^{-1}$, below which the measured values are unreliable as a result of the medium resolution of the spectrograph (see Sect. \ref{sec4}).
}
\label{SpecPars}
\end{figure}

To explain this phenomenon, we plot in Fig. \ref{SpecPars} the spectral parameters of the observed candidates, which were derived in Sect. \ref{sec4}. Since spectral type is a proxy of $T_{\rm eff}$, we place $T_{\rm eff}$ at the abscissa. Similarly to Fig. \ref{EXODAT}, there is an excess of cool FAs ($T_{\rm eff}\lesssim5600$\,K), with just a few confirmed binaries in that temperature regime. In addition, cool FAs seem to constitute a distinct sample of slowly rotating stars with lower gravity and lower metallicity. Even though our spectral-parameter measurement technique is prone to systematic biases \citep[e.g.,][]{Torres2012}, such a strong bimodality suggests that most cool FAs are possibly red giant stars.

We propose that the main reason that red giants introduce false candidates with photometric periods of $\gtrsim1.4$\,day is related to solar-like oscillations. Using \textit{Kepler} data, \citet{Mosser2013} have shown that solar-like oscillations of $1$--$2$\,$\rm{M}_{\odot}$ red giants have frequencies of $1$--$10$\,$\mu\rm{Hz}$ (periods of $1.2$--$12$\,days), amplitudes ($\rm{A_{max}}$) of $0.1$--$1$\,mmag, and that the amplitude increases with the period as a power law. For BEER binaries, the photometric amplitude decreases with the period \citep*{zma2007}. However, it appears that for periods of about $1$--$10$\,days, these two phenomena might have similar photometric amplitudes. A large fraction of the stars observed by CoRoT are indeed giants, particularly in the CoRoT-center fields \citep[e.g.,][]{ExoDatDeleuil2009,Gazzano2010}. Although solar-like oscillations are semi-regular in nature, given the typical length and S/N of long-run CoRoT data, BEER might have interpreted the variability of some red giants as an indication for the presence of a short-period companion.

To check whether filtering out red candidates with periods of $>1.4$\,days improves BEER's performance, we list in the right-hand side of Table \ref{tab:pri} the number of binary confirmations and FAs per priority class, after removing candidates with BEER-model $M_2\sin i<0.25 M_{\odot}$, and also candidates of spectral type later than G$7$ with photometric periods of $>1.4$\,days. These two filters applied to the candidate list lowers the fraction of FAs to $\sim1/3$ for priority $1$ candidates, and to $\sim2/3$ for priority $2$ candidates.

\subsection{BEER-model score}

\begin{figure}
\resizebox{\hsize}{!}
{\includegraphics{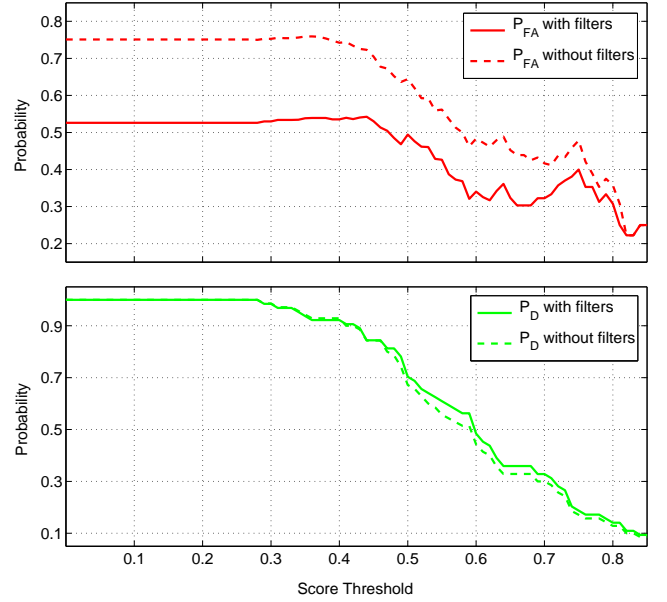}}
\caption{Upper panel: False-alarm probability as a function of the BEER score threshold. Lower panel: The detection probability as a function of the BEER score threshold. The dashed lines show the probabilities for the whole sample of $281$ candidates observed, while the solid lines show the probabilities for the remaining $135$ candidates, after applying the three filters discussed in the text (see also the first two rows of the right-hand side of Table \ref{tab:pri}). The plot was truncated at a score threshold of $0.85$ since only $8$ candidates got a higher score.}
\label{fig:roc}
\end{figure}

To evaluate the BEER performance as a function of its score, we counted the targets with scores higher than some threshold T and obtained the detection probability $P_D(T)$ and the false-alarm probability $P_{FA}(T) $ of the algorithm as
\begin{equation}
P_D = \frac{Number\, of\, confirmations\, with\, score>T}{Total\, number\, of\, confirmations}
\end{equation}
\begin{equation}
P_{FA} = \frac{Number\, of\, false\, alarms\, with\, score>T}{Total\, number\, of\, targets\, with\, score>T}\, .
\end{equation}

To illustrate the algorithm performance, Fig. \ref{fig:roc} plots $P_D$ and $P_{FA}$ as a function of T for two sets of targets---the whole sample of $281$ candidates observed, and the remaining $135$ candidates, after filtering out priority $3$ candidates, candidates with BEER-model $M_2\sin i<0.25 M_{\odot}$, and candidates of spectral type later than G$7$ with photometric periods of $>1.4$\,day. The addition of these three filters lowers $P_{FA}$ and raises $P_D$ for any given T, hence improves the algorithm performance.

Estimating the two probabilities for any selected threshold level might be useful in estimating the results of future CoRoT-based BEER searches. If such a search would yield $N$ targets with scores higher than some predefined threshold T, then using the $P_D$ and $P_{FA}$ that correspond to T, we expect to have
\begin{equation}
N_C = N (1-P_{FA}) \, ,
\end{equation}
\begin{equation}
N_{FA} = N P_{FA} \, , \rm{and}
\end{equation}
\begin{equation}
N_{CC} = \frac{N (1-P_{FA})}{P_D} \, ,
\end{equation}
where $N_C$ is the number of expected confirmations within the N candidates, $N_{FA}$ is the number of expected FAs within the N candidates, and $N_{CC}$ is the estimated number of binaries that can be discovered by BEER in the original sample of CoRoT targets. For instance, using the solid lines in Fig. \ref{fig:roc}, for $T=0.6$ we get $P_{FA}\sim1/3$ and $P_D\sim1/2$. This means that, for $T=0.6$, we expect about two thirds of the selected sample to be true BEER binaries and the total number of BEER binaries in the original sample of CoRoT targets to be about $(4/3) N$.

\section{Mass ratio and orbital period distribution of the BEER CoRoT sample}
\label{sec7}

\begin{table}
\caption{Spectral parameters, primary mass, and the assigned mass ratio of the $45$ confirmed BEER SB1s.}
\begin{tabular}{lccccc}
\hline
\hline
CoRoT & $T_{\rm eff}$ & log\,$g$ & $[\rm{m/H}]$ & Primary mass & Mass ratio\\
ID & (K) & (cgs) & (dex) & ($\rm{M}_{\odot}$) & ($\rm{M}_2/\rm{M}_1$) \\
\hline
$103924393$ & $6500$ & $4.6$ & $-0.1$ & $1.20 \pm 0.16$ & $0.69^{+0.70}_{-0.12}$ \\
$105767195$ & $6400$ & $4.6$ & $-0.4$ & $1.07 \pm 0.13$ & $0.28^{+0.22}_{-0.04}$ \\
$105810223$ & $6500$ & $3.8$ & $0.0$ & $1.59 \pm 0.38$ & $0.62^{+0.61}_{-0.11}$ \\
$104674562$ & $6500$ & $4.6$ & $-0.2$ & $1.15 \pm 0.15$ & $0.32^{+0.25}_{-0.05}$ \\
$310222016$ & $6800$ & $4.5$ & $-0.1$ & $1.28 \pm 0.17$ & $0.21^{+0.16}_{-0.03}$ \\
$310210722$ & $7800$ & $3.9$ & $0.1$ & $2.02 \pm 0.46$ & $0.52^{+0.48}_{-0.09}$ \\
$105818861$ & $6400$ & $4.6$ & $-0.2$ & $1.12 \pm 0.15$ & $0.41^{+0.36}_{-0.06}$ \\
$101029997$ & $6600$ & $4.6$ & $-0.1$ & $1.20 \pm 0.15$ & $0.41^{+0.35}_{-0.06}$ \\
$105844488$ & $5000$ & $5.5$ & $+0.3$ & $1.01 \pm 0.23$ & $0.53^{+0.50}_{-0.09}$ \\
$310215106$ & $6300$ & $3.9$ & $-0.2$ & $1.40 \pm 0.32$ & $0.28^{+0.23}_{-0.05}$ \\
$310176634$ & $5000$ & $5.5$ & $+0.5$ & $1.07 \pm 0.26$ & $0.10^{+0.07}_{-0.02}$ \\
$105336757$ & $6500$ & $4.6$ & $-0.2$ & $1.16 \pm 0.15$ & $0.26^{+0.20}_{-0.04}$ \\
$105712106$ & $6300$ & $4.6$ & $-0.0$ & $1.15 \pm 0.15$ & $0.48^{+0.43}_{-0.08}$ \\
$105403147$ & $6700$ & $4.5$ & $-0.2$ & $1.24 \pm 0.17$ & $0.30^{+0.24}_{-0.05}$ \\
$105706604$ & $6300$ & $4.5$ & $-0.1$ & $1.15 \pm 0.16$ & $0.26^{+0.20}_{-0.04}$ \\
$105714214$ & $6200$ & $5.1$ & $0.2$ & $1.19 \pm 0.17$ & $0.33^{+0.27}_{-0.05}$ \\
$104295292$ & $6200$ & $4.7$ & $-0.0$ & $1.12 \pm 0.14$ & $0.22^{+0.17}_{-0.03}$ \\
$110567660$ & $6500$ & $4.5$ & $0.1$ & $1.24 \pm 0.17$ & $0.32^{+0.26}_{-0.05}$ \\
$101177265$ & $6700$ & $4.2$ & $0.0$ & $1.42 \pm 0.27$ & $0.17^{+0.12}_{-0.03}$ \\
$101014035$ & $6100$ & $4.2$ & $-0.0$ & $1.24 \pm 0.24$ & $0.25^{+0.19}_{-0.04}$ \\
$100889978$ & $6300$ & $5.0$ & $-0.2$ & $1.11 \pm 0.14$ & $0.39^{+0.33}_{-0.06}$ \\
$105661774$ & $8900$ & $4.0$ & $-0.5$ & $1.96 \pm 0.40$ & $0.20^{+0.15}_{-0.04}$ \\
$105618890$ & $6300$ & $4.4$ & $0.0$ & $1.20 \pm 0.18$ & $0.34^{+0.28}_{-0.05}$ \\
$105802223$ & $6400$ & $4.6$ & $0.0$ & $1.20 \pm 0.16$ & $0.11^{+0.08}_{-0.02}$ \\
$100880613$ & $7300$ & $4.5$ & $+0.4$ & $1.60 \pm 0.22$ & $0.18^{+0.13}_{-0.03}$ \\
$105760939$ & $6200$ & $4.2$ & $0.1$ & $1.28 \pm 0.23$ & $0.14^{+0.10}_{-0.02}$ \\
$104113878$ & $6700$ & $4.7$ & $0.1$ & $1.27 \pm 0.15$ & $0.11^{+0.07}_{-0.02}$ \\
$103782315$ & $5900$ & $4.6$ & $+0.3$ & $1.12 \pm 0.15$ & $0.16^{+0.11}_{-0.02}$ \\
$310169750$ & $6500$ & $3.6$ & $+0.4$ & $1.96 \pm 0.52$ & $0.17^{+0.12}_{-0.03}$ \\
$103922738$ & $6000$ & $4.7$ & $0.1$ & $1.09 \pm 0.14$ & $0.07^{+0.05}_{-0.01}$ \\
$104648865$ & $8700$ & $3.3$ & $-0.4$ & $2.89 \pm 0.83$ & $0.15^{+0.12}_{-0.06}$ \\
$104536524$ & $6400$ & $4.4$ & $0.1$ & $1.25 \pm 0.20$ & $0.08^{+0.06}_{-0.01}$ \\
$101058035$ & $6500$ & $4.7$ & $0.1$ & $1.23 \pm 0.15$ & $0.11^{+0.07}_{-0.02}$ \\
$105378453$ & $6300$ & $5.0$ & $-0.1$ & $1.13 \pm 0.14$ & $0.17^{+0.12}_{-0.02}$ \\
$105597526$ & $6200$ & $3.0$ & $-0.0$ & $2.47 \pm 0.78$ & $0.17^{+0.13}_{-0.04}$ \\
$104667709$ & $6200$ & $3.6$ & $-0.0$ & $1.70 \pm 0.46$ & $0.11^{+0.07}_{-0.02}$ \\
$105154613$ & $6300$ & $4.8$ & $-0.2$ & $1.08 \pm 0.13$ & $0.09^{+0.06}_{-0.01}$ \\
$100537909$ & $6500$ & $5.0$ & $-0.1$ & $1.19 \pm 0.15$ & $0.24^{+0.19}_{-0.04}$ \\
$105659320$ & $6500$ & $4.6$ & $-0.1$ & $1.16 \pm 0.15$ & $0.03^{+0.02}_{-0.01}$ \\
$101044188$ & $5900$ & $4.8$ & $0.1$ & $1.07 \pm 0.13$ & $0.05^{+0.03}_{-0.01}$ \\
$104279119$ & $6200$ & $5.1$ & $+0.3$ & $1.24 \pm 0.18$ & $0.18^{+0.13}_{-0.03}$ \\
$104598628$ & $6500$ & $3.7$ & $0.2$ & $1.79 \pm 0.46$ & $0.13^{+0.09}_{-0.03}$ \\
$105164611$ & $8500$ & $4.5$ & $-0.1$ & $1.75 \pm 0.23$ & $0.18^{+0.13}_{-0.04}$ \\
$100851348$ & $6300$ & $3.9$ & $+0.3$ & $1.58 \pm 0.36$ & $0.06^{+0.04}_{-0.02}$ \\
$105472536$ & $5500$ & $3.9$ & $-0.0$ & $1.20 \pm 0.28$ & $0.14^{+0.10}_{-0.03}$ \\
\hline
\end{tabular}
\label{tab:s1}
\end{table}

\begin{table}
\caption{Spectral parameters, primary mass, and mass ratio of the $18$ confirmed BEER SB2s.}
\begin{tabular}{lccccc}
\hline
\hline
CoRoT & $T_{\rm eff}$ & log\,$g$ & $[\rm{m/H}]$ & Primary mass & Mass ratio\\
ID & (K) & (cgs) & (dex) & ($\rm{M}_{\odot}$) & ($\rm{M}_2/\rm{M}_1$) \\
\hline
$105962436$ & $8200$ & $4.4$ & $-0.2$ & $1.68 \pm 0.26$ & $0.62 \pm  0.04$ \\
$106024478$ & $6000$ & $3.2$ & $-0.1$ & $2.04 \pm 0.63$ & $0.99 \pm  0.02$ \\
$105649738$ & $5800$ & $3.1$ & $+0.3$ & $2.26 \pm 0.71$ & $1.00 \pm 0.02$ \\
$105928477$ & $6300$ & $4.4$ & $-0.0$ & $1.19 \pm 0.18$ & $0.99 \pm  0.07$ \\
$100688131$ & $6100$ & $4.7$ & $0.1$ & $1.14 \pm 0.14$ & $0.88 \pm  0.02$ \\
$310198235$ & $6500$ & $3.9$ & $+0.3$ & $1.66 \pm 0.39$ & $0.88 \pm 0.12$ \\
$100976101$ & $7400$ & $4.7$ & $0.1$ & $1.48 \pm 0.16$ & $0.60 \pm  0.04$ \\
$104369937$ & $6400$ & $5.0$ & $0.1$ & $1.18 \pm 0.14$ & $0.77 \pm  0.03$ \\
$105963904$ & $6100$ & $4.7$ & $0.1$ & $1.11 \pm 0.14$ & $0.82 \pm  0.02$ \\
$104181232$ & $6500$ & $5.0$ & $-0.0$ & $1.19 \pm 0.15$ & $0.70 \pm  0.10$ \\
$105506915$ & $6500$ & $3.8$ & $0.0$ & $1.62 \pm 0.40$ & $0.21 \pm  0.04$ \\
$310136399$ & $5900$ & $3.3$ & $+0.5$ & $2.04 \pm 0.60$ & $0.96 \pm 0.06$ \\
$101177998$ & $7200$ & $4.3$ & $-0.0$ & $1.50 \pm 0.24$ & $0.17 \pm  0.03$ \\
$310212616$ & $6100$ & $4.5$ & $+0.3$ & $1.22 \pm 0.17$ & $0.65 \pm 0.11$ \\
$103838038$ & $6300$ & $4.5$ & $-0.1$ & $1.13 \pm 0.16$ & $0.98 \pm  0.03$ \\
$310173237$ & $5200$ & $4.5$ & $0.2$ & $0.92 \pm 0.14$ & $0.65 \pm  0.15$ \\
$104432741$ & $5800$ & $2.9$ & $+0.4$ & $2.74 \pm 0.90$ & $0.52 \pm 0.07$ \\
$105583867$ & $7300$ & $3.8$ & $0.0$ & $1.92 \pm 0.47$ & $0.38 \pm  0.14$ \\
\hline
\end{tabular}
\label{tab:s2}
\end{table}

To discuss the mass ratio and period distribution of the sample of confirmed BEER binaries, we wish to plot an estimate of the mass ratio of each system as a function of its orbital period. For SB2s this is straightforward, since the mass ratio ($q\equiv M_2/M_1=K_1/K_2$) can be measured directly from the orbital parameters. For SB1s, however, the mass ratio depends not only on the orbital parameters, but also on the primary mass and orbital inclination, and can be found using the relation
\begin{equation}
(M_1f^{-1}\sin^3i)q^3-q^2-2q-1=0 $,$
\end{equation}
where $M_1$ is the estimated mass of the primary, $f$ is the mass function derived analytically from the orbital parameters, and $i$ is the inclination.

To estimate $M_1$ of the confirmed binaries, we used the empiric relations given in \citet{Torres2010}, which express the stellar mass and radius in terms of its observed spectral parameters. As an input, we used the spectral parameters that were derived for the primary component of the confirmed SB1s and SB2s in Sect. \ref{sec4}. To estimate the mass uncertainties, we took equal uncertainties of $300$\,K in $T_{\rm eff}$, $0.4$\,dex in log\,$g$, and $0.3$\,dex in $[\rm{m/H}]$ to all primary stars. For most of the observed stars these uncertainties are larger than the scatter of the best spectral parameters between consecutive exposures (see for instance Fig. \ref{Fig2_2}), but taking into account the possible systematic errors \citep[e.g.,][]{Torres2012}, they are probably reasonable. The intrinsic scatter from the empiric relations \citep{Torres2010} was added in quadrature to the mass uncertainties.

To assign an inclination for each SB1 we considered three options---to use the inclination estimated from its light curve by BEER, to derive its inclination distribution using the algorithm developed by \citet{MG1992a}, or to assume an isotropic inclination distribution. While using the BEER-model $\sin i$ estimates could sound appealing, we decided not to use them because we do not know their actual uncertainties, including possible systematic biases. Deriving inclination distributions using the algorithm of \citet{MG1992a} would probably be the correct way to proceed, but it is beyond the scope of this paper. We therefore, somewhat  arbitrarily,  assigned a value of $\sin i = 0.866^{+0.121}_{-0.326}$, taking the median of an isotropic inclination distribution and the confidence limits to cover the central $68.3\%$ of the distribution \citep[e.g.,][]{HoTurner2011,LopezJenkins2012}.

Table \ref{tab:s1} lists the spectral parameters, primary mass, and the assigned mass ratio of the confirmed BEER SB1s. Table \ref{tab:s2} lists the primary spectral parameters, primary mass, and the estimated mass ratio of the confirmed BEER SB2s. For convenience, the order of Tables \ref{tab:s1} and \ref{tab:s2} is the same as of Tables \ref{t3e} and \ref{t3f}. We did not estimate the masses of the seven diluted BEER binaries since their spectral parameters might have been biased by the presence of the third star.

\begin{figure}
\resizebox{\hsize}{!}
{\includegraphics{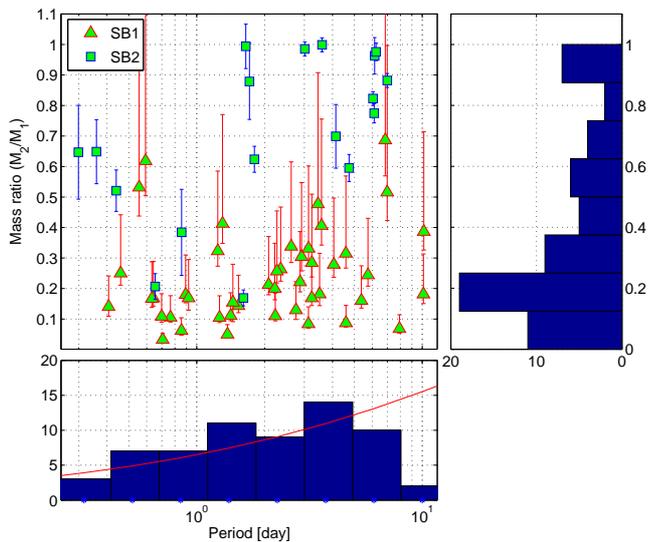}}
\caption{Assigned mass ratio of the $45$ confirmed SB1s and estimated mass ratio of the $18$ confirmed SB2s versus orbital period. For the SB1s we assumed an isotropic inclination distribution (see text). Bottom panel: period histogram. The solid red line shows the log-normal fit reported by \citet{Raghavan2010} to the period distribution of stellar binaries, scaled to best fit the histogram. Right panel: mass-ratio histogram.}
\label{qP}
\end{figure}

Figure \ref{qP} shows the assigned mass ratio of the $45$ confirmed SB1s and the estimated mass ratio of the $18$ confirmed SB2s  as a function of their orbital period. The period and the mass-ratio histograms are plotted as well. The solid red line in the period histogram shows the log-normal fit of \citet{Raghavan2010} to the period distribution of stellar binaries in the solar neighborhood, scaled to best fit the histogram.

One feature in Figure \ref{qP} is that for periods of $>1$\,day, there seems to be a transition from SB1s to SB2s at $q\sim0.6$. A similar transition was pointed out by \citet{Halbwachs2003} in their unbiased sample of solar-like spectroscopic binaries. In our sample this transition can be explained by the mass-luminosity relation of normal stars that is expected to be $L\sim M^4$ \citep[e.g.,][p. 382]{Torres2010,Cox2000}. Since the lowest significant $\alpha$ we could detect was $\sim0.1$ (see Table \ref{t3b}), we did not expect detached binaries with $q\lesssim0.6$ to be classified as SB2s. For periods of $\lesssim1$\,day this transition is less evident, since at such short periods some of these binaries could experience mass transfer via Roche-lobe overflow \citep[e.g.,][]{Eggleton1983}, and the mass-luminosity relation deviates from that of isolated normal stars \citep[e.g.,][p. 154]{Batten1973book}.

The period histogram clearly shows a rise in the number of binaries per $\Delta$log$P$ from $\sim0.3$ to $\sim6$\,days, followed by a sharp drop at periods of $>8$\,days. Since the detection probability of BEER binaries decreases with orbital period (see Fig. \ref{fig:m2sini}), this rise is probably real. In the past three decades, several studies reported a log-normal period distribution of solar-like binaries, with a peak at log$P\sim5$ \citep[e.g.,][]{DM1991,Raghavan2010}, and Fig. \ref{qP} shows that our period histogram fits well the log-normal fit of \citet{Raghavan2010} out to $P\sim8$\,days. The scaling factor between the two samples is $1.61$, which means that for a bin size of $\Delta$log$P=0.214$, as we use here, we would obtain $\sim74$ binaries at the peak of the distribution at log$P=5.03$. The histogram sharply drops at periods of $>8$\,days both because at such periods BEER approaches its sensitivity limit when applied to CoRoT long-run light curves and because the AAOmega observing run was limited to seven nights.

The mass-ratio distribution of short-period binaries was the subject of several in-depth studies in the last three decades \citep[e.g.,][]{Halbwachs1987,DM1991,MG1992b,Halbwachs2003,GML2003,Fisher2005}.
The main debates regarding the mass-ratio distribution of spectroscopic binaries are about the existence of peaks at $q\sim1$ and/or at $q\sim0.2$ and about the shape of the distribution toward lower $q$ values---whether it is monotonically increasing, decreasing, or flat. It is therefore of interest to plot the distribution of the assigned mass ratios of our sample, as was done in Fig.~\ref{qP}.

The histogram presents three features, which we describe from top to bottom. (1) A clear peak at $q\sim1$. (2) The number of binaries increases with decreasing $q$. (3) The histogram peaks at $q\sim0.2$ and then falls sharply towards $q\sim0.1$. However, correcting our sample for the \citet{Opik1924} effect might have significantly diminished the $q\sim1$ peak \citep[e.g.,][]{GML2003,Halbwachs2003}. In addition, at $q<0.6$ our histogram is dominated by SB1s, and \citet{MG1992a} have already shown that using an average $\sin i$ value to all SB1s, like we did here, might produce a monotonically increasing $q$ distribution when the actual true distribution is flat. Lastly, the drop toward $q\sim0.1$ can be explained both by BEER's sensitivity for CoRoT long-run light curves, which probably approaches its limit at such low mass-ratios, and by the limited RV precision of our AAOmega data. Therefore, before reaching any definite conclusion, our data need an in-depth analysis to derive a real mass-ratio distribution of the CoRoT BEER sample (Shahaf et al., in prep.). 

\section{Summary}
\label{sec8}

We have presented AAOmega RV confirmation of seventy new non-eclipsing short-period binaries found by BEER in CoRoT light curves. The confirmed binaries span a range of $0.3$--$10$\,days in orbital period, showing a clear rise in the number of binaries per $\Delta$log$P$ toward longer periods. Our spectral analysis shows that the primary stars in the detected binaries are typically of spectral type G or earlier, and the mass ratio (including the mass ratio assigned to SB1s) spans a range of $0.03$--$1$. The mass-ratio histogram resembles a double-peak distribution \citep[e.g.,][]{Halbwachs2003}, but since we did not correct our results, neither for some well-known selection effects \citep[e.g.,][]{Opik1924} nor for the non-isotropic inclination distribution \citep[e.g.,][]{MG1992a}, the underlying true distribution might also be flat. Nevertheless, the orbital solutions presented here can be used in the future for an in-depth study of the mass-ratio distribution of short-period binaries, similar to the study presented by \citet*{GML2003}.

On the lower end of our detection capability, we have detected two BD candidates on a $\sim1$\, day period orbit around F-G stars. We considered them as BD candidates and not bona fide BDs for two reasons: (1) the true inclination is not known, and (2) higher resolution and/or S/N spectroscopic observations might reveal a faint companion that could not have been found in the AAOmega spectra.

Relativistic beaming was already detected in CoRoT light curves \citep[e.g.,][]{Mazeh2010}, but this is the first time that BEER detected non-eclipsing binaries in CoRoT light
curves. To roughly estimate the expected number of such binaries that can potentially be detected with BEER in CoRoT long-run light curves we first need to correct for the $200$ binaries that were selected for follow-up, but were not observed, taking into account their BEER priorities, and then for the total number of CoRoT long-run light curves ($\sim110,000$), assuming that short-period binaries do not have a preference for center or anticenter fields. Doing so, we estimate that observing all $481$ selected candidates would have brought the number of confirmed binaries to $\sim110$, and that the expected number of beaming binaries that can be detected in CoRoT long-run light curves is $\sim300$. This confirms the prediction made by \citet*{zma2007}: CoRoT and \textit{Kepler} will be capable of detecting hundreds of non-eclipsing beaming binaries.

Investigating BEER's false-alarm probability and nature in CoRoT long-run light curves, we showed that red giants introduce a major source of false candidates and demonstrated a way to improve BEER's performance in extracting higher fidelity samples from future searches. Understanding the dependence of the fidelity and completeness of such future samples on parameters such as period, spectral type, and BEER-model score might enable using these samples to derive some statistical properties of the short-period binary population, like the period distribution, even before RV follow-up is done. Merging well-corrected large samples of CoRoT and \textit{Kepler} beaming binaries with equivalent EB samples \citep[e.g.,][]{Prsa2011AJ} might create large and well-defined samples of short-period binaries, which may shed some light on binary formation and evolution.

Radial-velocity follow-up observations of beaming and eclipsing binaries will continue to play a key role in these efforts because in most cases this is the only way to measure their mass ratio and the mass of the secondary companion. Scaling the AAOmega time that was required to make the detections reported here, we estimate that only three additional such runs would be required to detect virtually all beaming binaries that can be detected in CoRoT long-run light curves. This again demonstrates the efficiency of using multi-object spectrographs for this task \citep*[e.g.,][]{zma2007}. One advantage of the beaming binaries over the EBs is the ability of detecting binaries not only at edge-on inclinations, which widens the window for detecting intrinsically rare systems such as BD and massive-planetary companions to main-sequence stars \citep[e.g.,][]{Halbwachs2000, Faigler2013}.


\begin{acknowledgements}
We thank Gary Da Costa and Simon O'Toole for their help with preparing the AAOmega proposal and observations. We are grateful for the invaluable assistance of the AAOmega technical staff and support astronomers. Particularly, we thank Sarah Brough for her help with preparing, performing, and reducing the observations. We thank Gil Sokol for his help with the observations and the spectral analysis. We also thank Slavek M. Rucinski, the referee, for his fruitful comments and suggestions. The research leading to these results has received funding from the European Community's Seventh Framework Programme (FP7/2007-2013) under grant-agreement numbers 291352 (ERC) and RG226604 (OPTICON). This research was also supported by the Israel Science Foundation (grant No. 1423/11), and by the Israeli Centers of Research Excellence (I-CORE, grant No. 1829/12). The CoRoT space mission, launched on 2006 December 27, was developed and operated by the CNES, with participation of the Science Programs of ESA, ESA's RSSD, Austria, Belgium, Brazil, Germany, and Spain.
\end{acknowledgements}

\bibliographystyle{aa}
\bibliography{AAO.bib}

\begin{thebibliography}{65}
\expandafter\ifx\csname natexlab\endcsname\relax\def\natexlab#1{#1}\fi

\bibitem[{{Aigrain} {et~al.}(2008){Aigrain}, {Barge}, {Deleuil}, {Fressin},
  {Moutou}, {Queloz}, {Auvergne}, \& {Baglin}}]{Aigrain2008}
{Aigrain}, S., {Barge}, P., {Deleuil}, M., {et~al.} 2008, in Astronomical
  Society of the Pacific Conference Series, Vol. 384, 14th Cambridge Workshop
  on Cool Stars, Stellar Systems, and the Sun, 270--280

\bibitem[{{Aigrain} {et~al.}(2004){Aigrain}, {Favata}, \&
  {Gilmore}}]{aigrain04}
{Aigrain}, S., {Favata}, F., \& {Gilmore}, G. 2004, \aap, 414, 1139

\bibitem[{{Auvergne} {et~al.}(2009){Auvergne}, {Bodin}, {Boisnard}, {Buey},
  {Chaintreuil}, {Epstein}, {Jouret}, {Lam-Trong}, {Levacher}, {Magnan},
  {Perez}, {Plasson}, {Plesseria}, {Peter}, {Steller}, {Tiph{\`e}ne}, {Baglin},
  {Agogu{\'e}}, {Appourchaux}, {Barbet}, {Beaufort}, {Bellenger}, {Berlin},
  {Bernardi}, {Blouin}, {Boumier}, {Bonneau}, {Briet}, {Butler}, {Cautain},
  {Chiavassa}, {Costes}, {Cuvilho}, {Cunha-Parro}, {de Oliveira Fialho},
  {Decaudin}, {Defise}, {Djalal}, {Docclo}, {Drummond}, {Dupuis}, {Exil},
  {Faur{\'e}}, {Gaboriaud}, {Gamet}, {Gavalda}, {Grolleau}, {Gueguen},
  {Guivarc'h}, {Guterman}, {Hasiba}, {Huntzinger}, {Hustaix}, {Imbert},
  {Jeanville}, {Johlander}, {Jorda}, {Journoud}, {Karioty}, {Kerjean},
  {Lafond}, {Lapeyrere}, {Landiech}, {Larqu{\'e}}, {Laudet}, {Le Merrer},
  {Leporati}, {Leruyet}, {Levieuge}, {Llebaria}, {Martin}, {Mazy}, {Mesnager},
  {Michel}, {Moalic}, {Monjoin}, {Naudet}, {Neukirchner}, {Nguyen-Kim},
  {Ollivier}, {Orcesi}, {Ottacher}, {Oulali}, {Parisot}, {Perruchot},
  {Piacentino}, {Pinheiro da Silva}, {Platzer}, {Pontet}, {Pradines},
  {Quentin}, {Rohbeck}, {Rolland}, {Rollenhagen}, {Romagnan}, {Russ}, {Samadi},
  {Schmidt}, {Schwartz}, {Sebbag}, {Smit}, {Sunter}, {Tello}, {Toulouse},
  {Ulmer}, {Vandermarcq}, {Vergnault}, {Wallner}, {Waultier}, \&
  {Zanatta}}]{auvergne09}
{Auvergne}, M., {Bodin}, P., {Boisnard}, L., {et~al.} 2009, \aap, 506, 411

\bibitem[{{Baglin}(2003)}]{Baglin2003}
{Baglin}, A. 2003, Advances in Space Research, 31, 345

\bibitem[{{Batten}(1973)}]{Batten1973book}
{Batten}, A.~H. 1973, {Binary and multiple systems of stars}

\bibitem[{{Borucki} {et~al.}(2010){Borucki}, {Koch}, {Basri}, {Batalha},
  {Brown}, {Caldwell}, {Caldwell}, {Christensen-Dalsgaard}, {Cochran},
  {DeVore}, {Dunham}, {Dupree}, {Gautier}, {Geary}, {Gilliland}, {Gould},
  {Howell}, {Jenkins}, {Kondo}, {Latham}, {Marcy}, {Meibom}, {Kjeldsen},
  {Lissauer}, {Monet}, {Morrison}, {Sasselov}, {Tarter}, {Boss}, {Brownlee},
  {Owen}, {Buzasi}, {Charbonneau}, {Doyle}, {Fortney}, {Ford}, {Holman},
  {Seager}, {Steffen}, {Welsh}, {Rowe}, {Anderson}, {Buchhave}, {Ciardi},
  {Walkowicz}, {Sherry}, {Horch}, {Isaacson}, {Everett}, {Fischer}, {Torres},
  {Johnson}, {Endl}, {MacQueen}, {Bryson}, {Dotson}, {Haas}, {Kolodziejczak},
  {Van Cleve}, {Chandrasekaran}, {Twicken}, {Quintana}, {Clarke}, {Allen},
  {Li}, {Wu}, {Tenenbaum}, {Verner}, {Bruhweiler}, {Barnes}, \&
  {Prsa}}]{Borucki2010Sci}
{Borucki}, W.~J., {Koch}, D., {Basri}, G., {et~al.} 2010, Science, 327, 977

\bibitem[{{Carter} {et~al.}(2011){Carter}, {Rappaport}, \&
  {Fabrycky}}]{Carter2011ApJ}
{Carter}, J.~A., {Rappaport}, S., \& {Fabrycky}, D. 2011, \apj, 728, 139

\bibitem[{{Cox}(2000)}]{Cox2000}
{Cox}, A.~N. 2000, {Allen's astrophysical quantities}

\bibitem[{{Deleuil} {et~al.}(2009){Deleuil}, {Meunier}, {Moutou}, {Surace},
  {Deeg}, {Barbieri}, {Debosscher}, {Almenara}, {Agneray}, {Granet},
  {Guterman}, \& {Hodgkin}}]{ExoDatDeleuil2009}
{Deleuil}, M., {Meunier}, J.~C., {Moutou}, C., {et~al.} 2009, \aj, 138, 649

\bibitem[{{Deleuil} {et~al.}(2011){Deleuil}, {Moutou}, \&
  {Bord{\'e}}}]{Deleuil2011}
{Deleuil}, M., {Moutou}, C., \& {Bord{\'e}}, P. 2011, Detection and Dynamics of
  Transiting Exoplanets, St.~Michel l'Observatoire, France, Edited by
  F.~Bouchy; R.~D{\'{\i}}az; C.~Moutou; EPJ Web of Conferences, Volume 11,
  id.01001, 110, 1001

\bibitem[{{Duquennoy} \& {Mayor}(1991)}]{DM1991}
{Duquennoy}, A. \& {Mayor}, M. 1991, \aap, 248, 485

\bibitem[{{Eggleton}(1983)}]{Eggleton1983}
{Eggleton}, P.~P. 1983, \apj, 268, 368

\bibitem[{{Faigler} \& {Mazeh}(2011)}]{Faigler2011}
{Faigler}, S. \& {Mazeh}, T. 2011, \mnras, 415, 3921

\bibitem[{{Faigler} \& {Mazeh}(2015)}]{Faigler2015}
{Faigler}, S. \& {Mazeh}, T. 2015, \apj, 800, 73

\bibitem[{{Faigler} {et~al.}(2012){Faigler}, {Mazeh}, {Quinn}, {Latham}, \&
  {Tal-Or}}]{Faigler2012}
{Faigler}, S., {Mazeh}, T., {Quinn}, S.~N., {Latham}, D.~W., \& {Tal-Or}, L.
  2012, \apj, 746, 185

\bibitem[{{Faigler} {et~al.}(2013){Faigler}, {Tal-Or}, {Mazeh}, {Latham}, \&
  {Buchhave}}]{Faigler2013}
{Faigler}, S., {Tal-Or}, L., {Mazeh}, T., {Latham}, D.~W., \& {Buchhave}, L.~A.
  2013, \apj, 771, 26

\bibitem[{{Fisher} {et~al.}(2005){Fisher}, {Schr{\"o}der}, \&
  {Smith}}]{Fisher2005}
{Fisher}, J., {Schr{\"o}der}, K.-P., \& {Smith}, R.~C. 2005, \mnras, 361, 495

\bibitem[{{Gazzano} {et~al.}(2010){Gazzano}, {de Laverny}, {Deleuil},
  {Recio-Blanco}, {Bouchy}, {Moutou}, {Bijaoui}, {Ordenovic}, {Gandolfi}, \&
  {Loeillet}}]{Gazzano2010}
{Gazzano}, J.-C., {de Laverny}, P., {Deleuil}, M., {et~al.} 2010, \aap, 523,
  A91

\bibitem[{{Goldberg} {et~al.}(2003){Goldberg}, {Mazeh}, \& {Latham}}]{GML2003}
{Goldberg}, D., {Mazeh}, T., \& {Latham}, D.~W. 2003, \apj, 591, 397

\bibitem[{{Gray}(2005)}]{Gray2005}
{Gray}, D.~F. 2005, {The Observation and Analysis of Stellar Photospheres, 3rd
  Edition}, 465

\bibitem[{{Halbwachs}(1987)}]{Halbwachs1987}
{Halbwachs}, J.~L. 1987, \aap, 183, 234

\bibitem[{{Halbwachs} {et~al.}(2000){Halbwachs}, {Arenou}, {Mayor}, {Udry}, \&
  {Queloz}}]{Halbwachs2000}
{Halbwachs}, J.~L., {Arenou}, F., {Mayor}, M., {Udry}, S., \& {Queloz}, D.
  2000, \aap, 355, 581

\bibitem[{{Halbwachs} {et~al.}(2003){Halbwachs}, {Mayor}, {Udry}, \&
  {Arenou}}]{Halbwachs2003}
{Halbwachs}, J.~L., {Mayor}, M., {Udry}, S., \& {Arenou}, F. 2003, \aap, 397,
  159

\bibitem[{{Harrison} {et~al.}(2003){Harrison}, {Howell}, {Huber}, {Osborne},
  {Holtzman}, {Cash}, \& {Gelino}}]{Harrison2003}
{Harrison}, T.~E., {Howell}, S.~B., {Huber}, M.~E., {et~al.} 2003, \aj, 125,
  2609

\bibitem[{{Hauschildt} {et~al.}(1999){Hauschildt}, {Allard}, \&
  {Baron}}]{phoenix99}
{Hauschildt}, P.~H., {Allard}, F., \& {Baron}, E. 1999, \apj, 512, 377

\bibitem[{{Ho} \& {Turner}(2011)}]{HoTurner2011}
{Ho}, S. \& {Turner}, E.~L. 2011, \apj, 739, 26

\bibitem[{{Holland} \& {Welsch}(1977)}]{HW1997}
{Holland}, P.~W. \& {Welsch}, R.~E. 1977, Commun. Statist. -- Theor. Meth., 6,
  813

\bibitem[{{Jackson} {et~al.}(2012){Jackson}, {Lewis}, {Barnes}, {Drake Deming},
  {Showman}, \& {Fortney}}]{Jackson2012}
{Jackson}, B.~K., {Lewis}, N.~K., {Barnes}, J.~W., {et~al.} 2012, \apj, 751,
  112

\bibitem[{{Koch} {et~al.}(2010){Koch}, {Borucki}, {Basri}, {Batalha}, {Brown},
  {Caldwell}, {Christensen-Dalsgaard}, {Cochran}, {DeVore}, {Dunham},
  {Gautier}, {Geary}, {Gilliland}, {Gould}, {Jenkins}, {Kondo}, {Latham},
  {Lissauer}, {Marcy}, {Monet}, {Sasselov}, {Boss}, {Brownlee}, {Caldwell},
  {Dupree}, {Howell}, {Kjeldsen}, {Meibom}, {Morrison}, {Owen}, {Reitsema},
  {Tarter}, {Bryson}, {Dotson}, {Gazis}, {Haas}, {Kolodziejczak}, {Rowe}, {Van
  Cleve}, {Allen}, {Chandrasekaran}, {Clarke}, {Li}, {Quintana}, {Tenenbaum},
  {Twicken}, \& {Wu}}]{Koch2010ApJ}
{Koch}, D.~G., {Borucki}, W.~J., {Basri}, G., {et~al.} 2010, \apjl, 713, L79

\bibitem[{Kramida {et~al.}(2013)Kramida, {Yu.~Ralchenko}, Reader, \& {and NIST
  ASD Team}}]{NIST_ASD}
Kramida, A., {Yu.~Ralchenko}, Reader, J., \& {and NIST ASD Team}. 2013, {NIST
  Atomic Spectra Database (ver. 5.1), [Online]. Available:
  {\tt{http://physics.nist.gov/asd}} [2014, August 11]. National Institute of
  Standards and Technology, Gaithersburg, MD.}

\bibitem[{{Lane} {et~al.}(2011){Lane}, {Kiss}, {Lewis}, {Ibata}, {Siebert},
  {Bedding}, {Sz{\'e}kely}, \& {Szab{\'o}}}]{Lane2011}
{Lane}, R.~R., {Kiss}, L.~L., {Lewis}, G.~F., {et~al.} 2011, \aap, 530, A31

\bibitem[{{Lewis} {et~al.}(2002){Lewis}, {Cannon}, {Taylor}, {Glazebrook},
  {Bailey}, {Baldry}, {Barton}, {Bridges}, {Dalton}, {Farrell}, {Gray},
  {Lankshear}, {McCowage}, {Parry}, {Sharples}, {Shortridge}, {Smith},
  {Stevenson}, {Straede}, {Waller}, {Whittard}, {Wilcox}, \&
  {Willis}}]{Lewis2002a}
{Lewis}, I.~J., {Cannon}, R.~D., {Taylor}, K., {et~al.} 2002, \mnras, 333, 279

\bibitem[{{Loeb} \& {Gaudi}(2003)}]{Loeb2003}
{Loeb}, A. \& {Gaudi}, B.~S. 2003, \apjl, 588, L117

\bibitem[{{Lopez} \& {Jenkins}(2012)}]{LopezJenkins2012}
{Lopez}, S. \& {Jenkins}, J.~S. 2012, \apj, 756, 177

\bibitem[{{Lucy} \& {Sweeney}(1971)}]{Lucy1971}
{Lucy}, L.~B. \& {Sweeney}, M.~A. 1971, \aj, 76, 544

\bibitem[{{Mazeh}(2008)}]{mazeh08}
{Mazeh}, T. 2008, ArXiv e-prints, 801

\bibitem[{{Mazeh} \& {Faigler}(2010)}]{Mazeh2010}
{Mazeh}, T. \& {Faigler}, S. 2010, \aap, 521, L59+

\bibitem[{{Mazeh} \& {Goldberg}(1992)}]{MG1992a}
{Mazeh}, T. \& {Goldberg}, D. 1992, \apj, 394, 592

\bibitem[{{Mazeh} {et~al.}(1992){Mazeh}, {Goldberg}, {Duquennoy}, \&
  {Mayor}}]{MG1992b}
{Mazeh}, T., {Goldberg}, D., {Duquennoy}, A., \& {Mayor}, M. 1992, \apj, 401,
  265

\bibitem[{{Mazeh} {et~al.}(2012){Mazeh}, {Nachmani}, {Sokol}, {Faigler}, \&
  {Zucker}}]{Mazeh2012}
{Mazeh}, T., {Nachmani}, G., {Sokol}, G., {Faigler}, S., \& {Zucker}, S. 2012,
  \aap, 541, A56

\bibitem[{{Mazeh} \& {Zucker}(1994)}]{zmapss94}
{Mazeh}, T. \& {Zucker}, S. 1994, \apss, 212, 349

\bibitem[{{Mislis} {et~al.}(2012){Mislis}, {Heller}, {Schmitt}, \&
  {Hodgkin}}]{Mislis2012}
{Mislis}, D., {Heller}, R., {Schmitt}, J.~H.~M.~M., \& {Hodgkin}, S. 2012,
  \aap, 538, A4

\bibitem[{{Miszalski} {et~al.}(2006){Miszalski}, {Shortridge}, {Saunders},
  {Parker}, \& {Croom}}]{Miszalski2006}
{Miszalski}, B., {Shortridge}, K., {Saunders}, W., {Parker}, Q.~A., \& {Croom},
  S.~M. 2006, \mnras, 371, 1537

\bibitem[{{Morris}(1985)}]{Morris1985}
{Morris}, S.~L. 1985, \apj, 295, 143

\bibitem[{{Mosser} {et~al.}(2013){Mosser}, {Dziembowski}, {Belkacem}, {Goupil},
  {Michel}, {Samadi}, {Soszy{\'n}ski}, {Vrard}, {Elsworth}, {Hekker}, \&
  {Mathur}}]{Mosser2013}
{Mosser}, B., {Dziembowski}, W.~A., {Belkacem}, K., {et~al.} 2013, \aap, 559,
  A137

\bibitem[{{Moutou} {et~al.}(2013){Moutou}, {Deleuil}, {Guillot}, {Baglin},
  {Bord{\'e}}, {Bouchy}, {Cabrera}, {Csizmadia}, \& {Deeg}}]{Moutou2013}
{Moutou}, C., {Deleuil}, M., {Guillot}, T., {et~al.} 2013, \icarus, 226, 1625

\bibitem[{{{\"O}pik}(1924)}]{Opik1924}
{{\"O}pik}, E. 1924, Publications of the Tartu Astrofizica Observatory, 25, No
  6

\bibitem[{{Pr{\v s}a} {et~al.}(2011){Pr{\v s}a}, {Batalha}, {Slawson}, {Doyle},
  {Welsh}, {Orosz}, {Seager}, {Rucker}, {Mjaseth}, {Engle}, {Conroy},
  {Jenkins}, {Caldwell}, {Koch}, \& {Borucki}}]{Prsa2011AJ}
{Pr{\v s}a}, A., {Batalha}, N., {Slawson}, R.~W., {et~al.} 2011, \aj, 141, 83

\bibitem[{{Raghavan} {et~al.}(2010){Raghavan}, {McAlister}, {Henry}, {Latham},
  {Marcy}, {Mason}, {Gies}, {White}, \& {ten Brummelaar}}]{Raghavan2010}
{Raghavan}, D., {McAlister}, H.~A., {Henry}, T.~J., {et~al.} 2010, \apjs, 190,
  1

\bibitem[{{Rouan} {et~al.}(1998){Rouan}, {Baglin}, {Copet}, {Schneider},
  {Barge}, {Deleuil}, {Vuillemin}, \& {L{\'e}ger}}]{rouan98}
{Rouan}, D., {Baglin}, A., {Copet}, E., {et~al.} 1998, Earth Moon and Planets,
  81, 79

\bibitem[{{Rowe} {et~al.}(2014){Rowe}, {Bryson}, {Marcy}, {Lissauer},
  {Jontof-Hutter}, {Mullally}, {Gilliland}, {Issacson}, {Ford}, {Howell},
  {Borucki}, {Haas}, {Huber}, {Steffen}, {Thompson}, {Quintana}, {Barclay},
  {Still}, {Fortney}, {Gautier}, {Hunter}, {Caldwell}, {Ciardi}, {Devore},
  {Cochran}, {Jenkins}, {Agol}, {Carter}, \& {Geary}}]{Rowe2014}
{Rowe}, J.~F., {Bryson}, S.~T., {Marcy}, G.~W., {et~al.} 2014, \apj, 784, 45

\bibitem[{{Rybicki} \& {Lightman}(1979)}]{RL1979}
{Rybicki}, G.~B. \& {Lightman}, A.~P. 1979, {Radiative processes in
  astrophysics}

\bibitem[{{Santerne} {et~al.}(2012){Santerne}, {Moutou}, {Barros}, {Damiani},
  {D{\'{\i}}az}, {Almenara}, {Bonomo}, {Bouchy}, {Deleuil}, \&
  {H{\'e}brard}}]{Santerne2012KOI13}
{Santerne}, A., {Moutou}, C., {Barros}, S.~C.~C., {et~al.} 2012, \aap, 544, L12

\bibitem[{{Saunders} {et~al.}(2004){Saunders}, {Bridges}, {Gillingham},
  {Haynes}, {Smith}, {Whittard}, {Churilov}, {Lankshear}, {Croom}, {Jones}, \&
  {Boshuizen}}]{Saunders2004}
{Saunders}, W., {Bridges}, T., {Gillingham}, P., {et~al.} 2004, in SPIE
  Conference Series, ed. A.~F.~M. {Moorwood} \& M.~{Iye}, Vol. 5492, 389--400

\bibitem[{{Sebastian} {et~al.}(2012){Sebastian}, {Guenther}, {Schaffenroth},
  {Gandolfi}, {Geier}, {Heber}, {Deleuil}, \& {Moutou}}]{Sebastian2012}
{Sebastian}, D., {Guenther}, E.~W., {Schaffenroth}, V., {et~al.} 2012, \aap,
  541, A34

\bibitem[{{Shporer} {et~al.}(2011){Shporer}, {Jenkins}, {Rowe}, {Sanderfer},
  {Seader}, {Smith}, {Still}, {Thompson}, {Twicken}, \& {Welsh}}]{Shporer2011}
{Shporer}, A., {Jenkins}, J.~M., {Rowe}, J.~F., {et~al.} 2011, \aj, 142, 195

\bibitem[{{Smith} {et~al.}(2004){Smith}, {Saunders}, {Bridges}, {Churilov},
  {Lankshear}, {Dawson}, {Correll}, {Waller}, {Haynes}, \& {Frost}}]{Smith2004}
{Smith}, G.~A., {Saunders}, W., {Bridges}, T., {et~al.} 2004, in SPIE
  Conference Series, ed. A.~F.~M. {Moorwood} \& M.~{Iye}, Vol. 5492, 410--420

\bibitem[{{Surace} {et~al.}(2008){Surace}, {Alonso}, {Barge}, {Cautain},
  {Chabaud}, {Deleuil}, {Fenouillet}, {Meunier}, \& {Moutou}}]{CoRoTalarm}
{Surace}, C., {Alonso}, R., {Barge}, P., {et~al.} 2008, in SPIE Conference
  Series, Vol. 7019

\bibitem[{{Taylor} {et~al.}(1996){Taylor}, {Bailey}, {Wilkins}, {Shortridge},
  \& {Glazebrook}}]{2DFDR}
{Taylor}, K., {Bailey}, J., {Wilkins}, T., {Shortridge}, K., \& {Glazebrook},
  K. 1996, in Astronomical Society of the Pacific Conference Series, Vol. 101,
  Astronomical Data Analysis Software and Systems V, ed. G.~H. {Jacoby} \&
  J.~{Barnes}, 195

\bibitem[{{Torres} {et~al.}(2010){Torres}, {Andersen}, \&
  {Gim{\'e}nez}}]{Torres2010}
{Torres}, G., {Andersen}, J., \& {Gim{\'e}nez}, A. 2010, \aapr, 18, 67

\bibitem[{{Torres} {et~al.}(2012){Torres}, {Fischer}, {Sozzetti}, {Buchhave},
  {Winn}, {Holman}, \& {Carter}}]{Torres2012}
{Torres}, G., {Fischer}, D.~A., {Sozzetti}, A., {et~al.} 2012, \apj, 757, 161

\bibitem[{{van Kerkwijk} {et~al.}(2010){van Kerkwijk}, {Rappaport}, {Breton},
  {Justham}, {Podsiadlowski}, \& {Han}}]{vanKerkwijk2010}
{van Kerkwijk}, M.~H., {Rappaport}, S.~A., {Breton}, R.~P., {et~al.} 2010,
  \apj, 715, 51

\bibitem[{{Wilson}(1990)}]{Wilson1990}
{Wilson}, R.~E. 1990, \apj, 356, 613

\bibitem[{{Zucker} \& {Mazeh}(1994)}]{zm94}
{Zucker}, S. \& {Mazeh}, T. 1994, \apj, 420, 806

\bibitem[{{Zucker} {et~al.}(2007){Zucker}, {Mazeh}, \& {Alexander}}]{zma2007}
{Zucker}, S., {Mazeh}, T., \& {Alexander}, T. 2007, \apj, 670, 1326

\end{thebibliography}


\begin{figure*}[!htp]
\resizebox{18.4cm}{!}
{\includegraphics{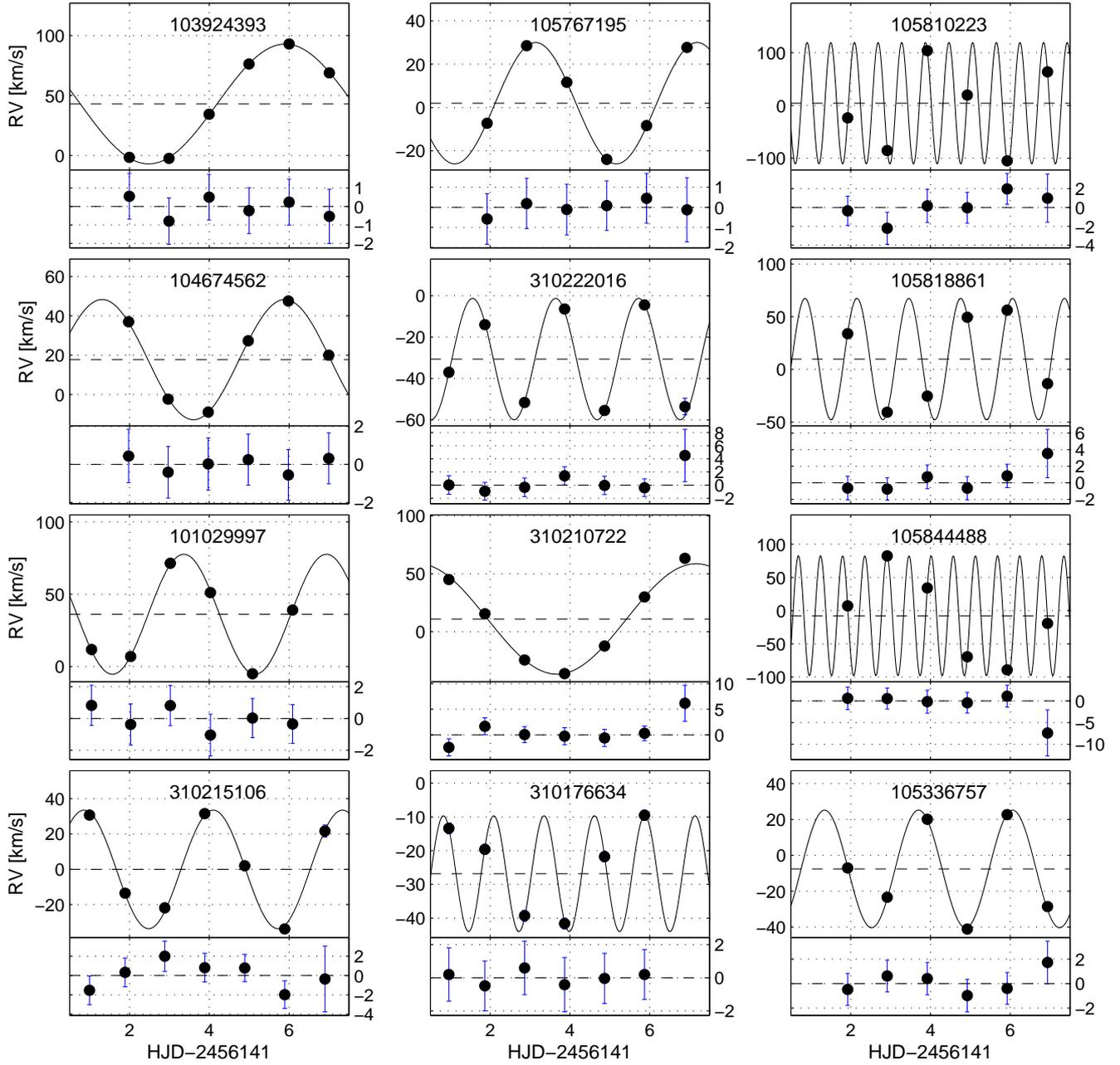}}
\caption{AAOmega RVs (black circles) and the best-fit Keplerian models (solid lines) of the confirmed BEER SB1s listed in Table \ref{t3e}.}
\label{Fig6}
\end{figure*}

\addtocounter{figure}{-1}

\begin{figure*}[!htp]
\resizebox{18.4cm}{!}
{\includegraphics{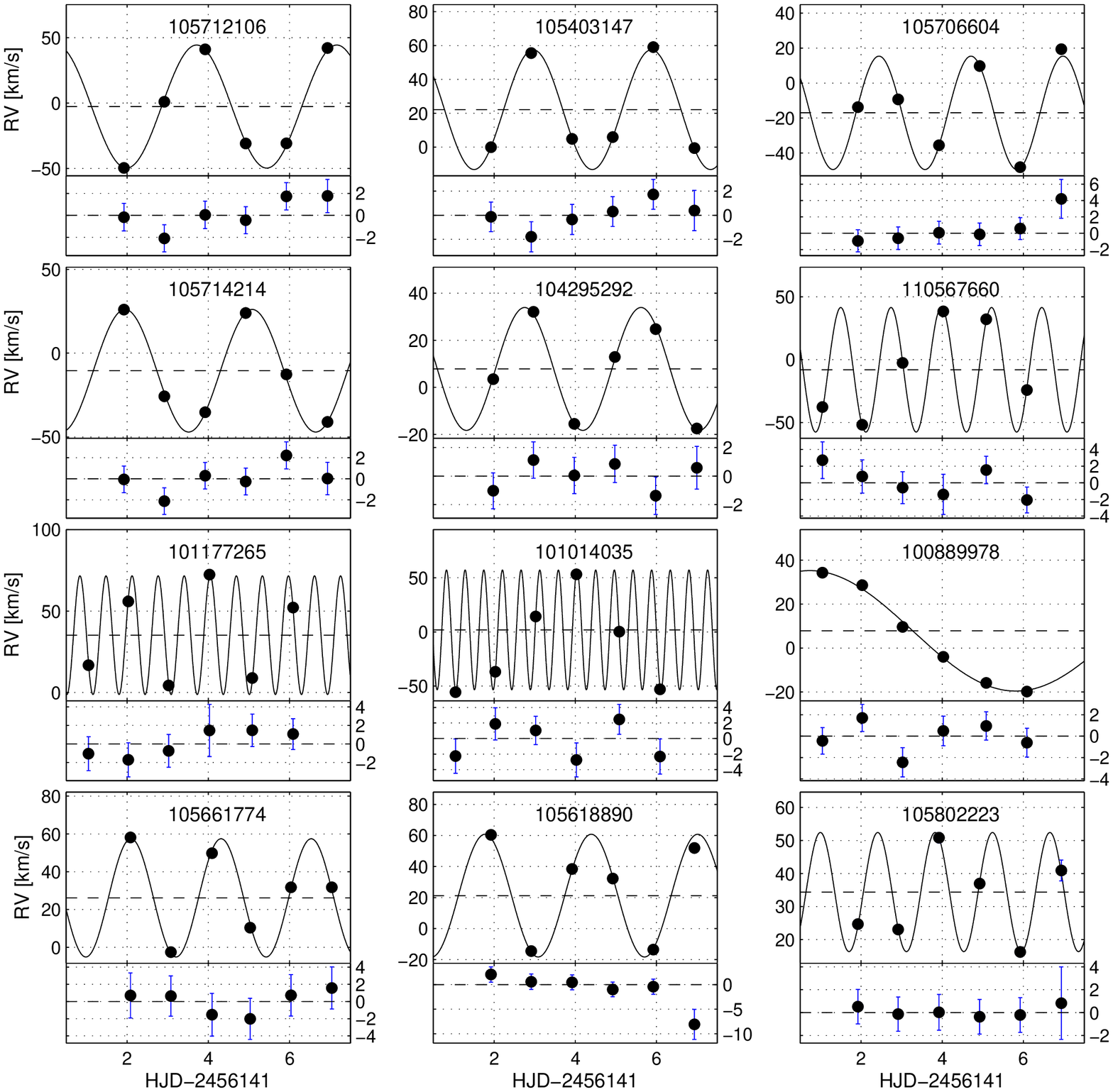}}
\caption{Continued.}
\label{Fig6}
\end{figure*}

\addtocounter{figure}{-1}

\begin{figure*}[!htp]
\resizebox{18.4cm}{!}
{\includegraphics{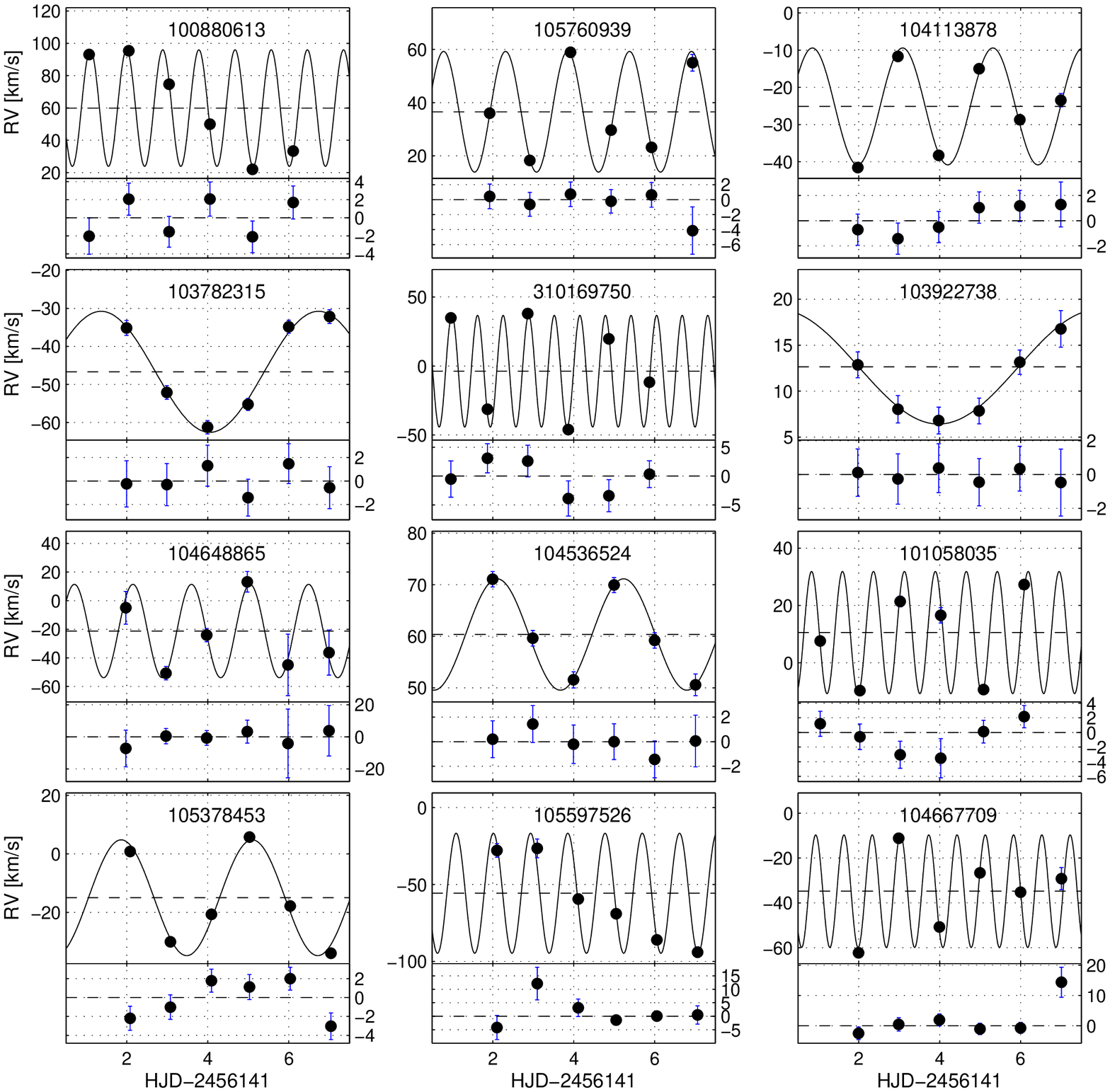}}
\caption{Continued.}
\label{Fig6}
\end{figure*}

\addtocounter{figure}{-1}

\begin{figure*}[!htp]
\resizebox{18.4cm}{!}
{\includegraphics{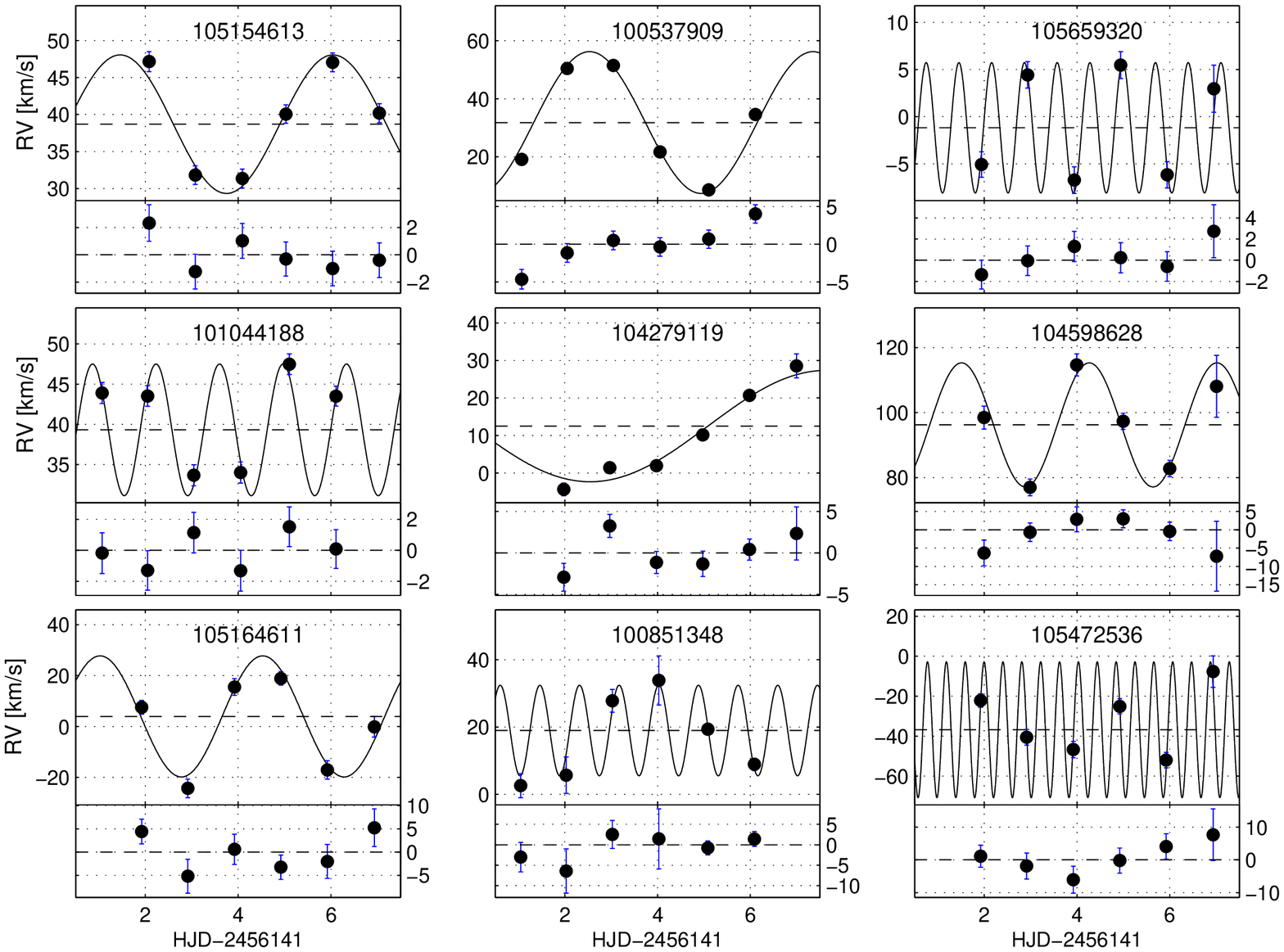}}
\caption{Continued.}
\label{Fig6}
\end{figure*}

\begin{figure*}[!htp]
\resizebox{18.4cm}{!}
{\includegraphics{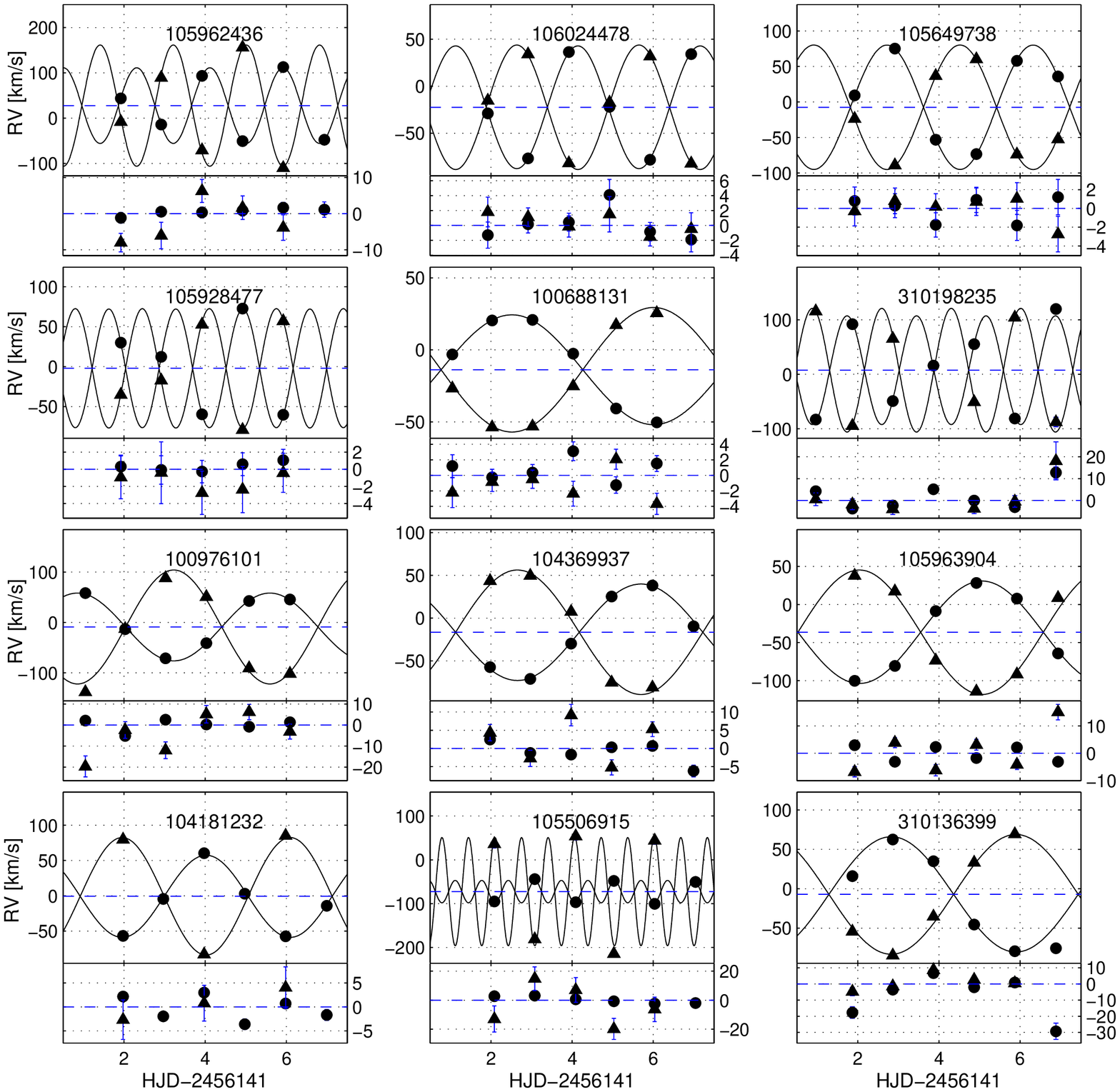}}
\caption{AAOmega RVs and the best-fit Keplerian models (solid lines) of the confirmed BEER SB2s listed in Table \ref{t3f}. Primary RVs are marked with circles and secondary RVs with triangles.}
\label{Fig7}
\end{figure*}

\addtocounter{figure}{-1}

\begin{figure*}[!htp]
\resizebox{18.4cm}{!}
{\includegraphics{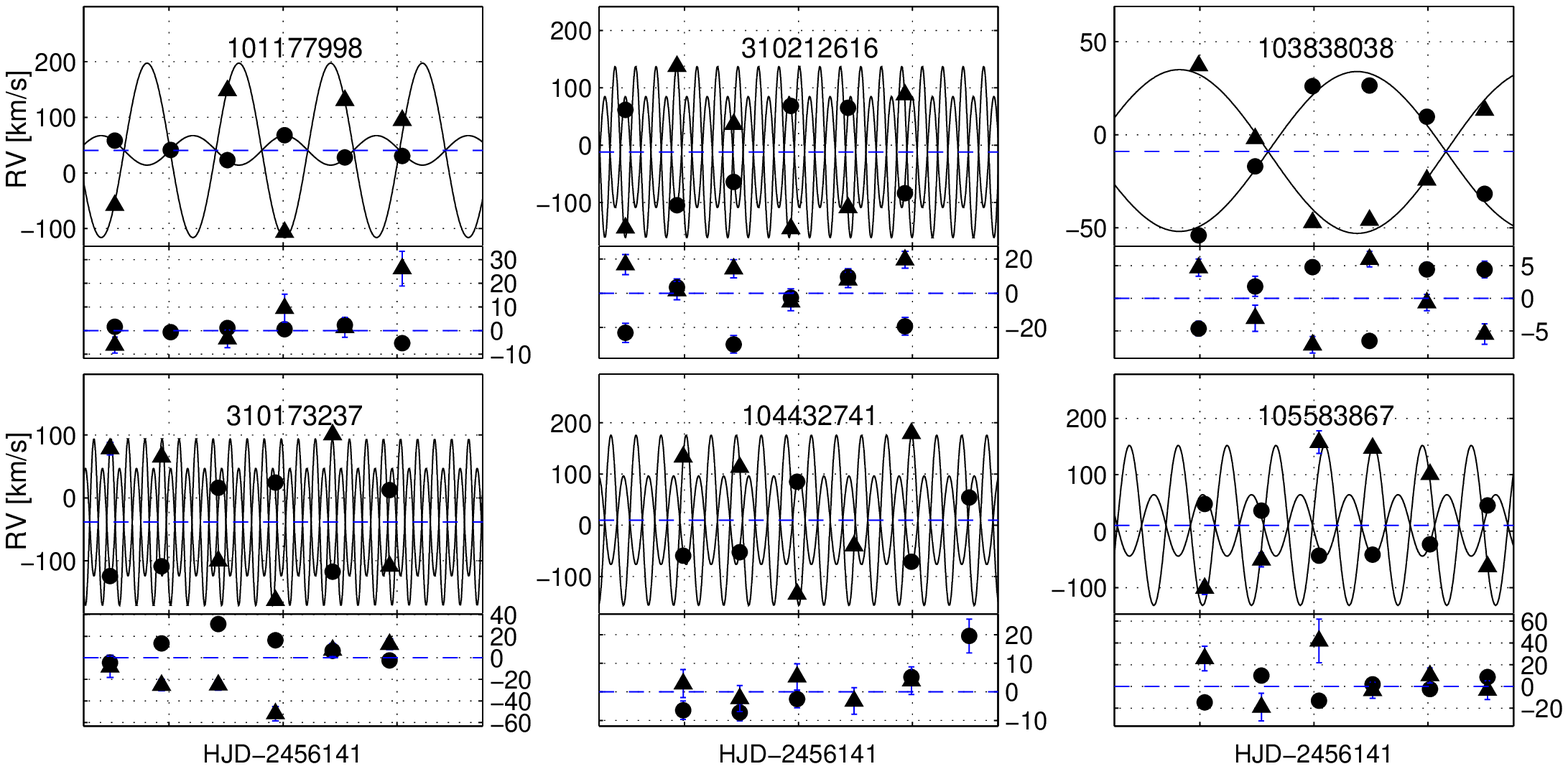}}
\caption{Continued.}
\label{Fig7}
\end{figure*}

\begin{figure*}[!htp]
\resizebox{18.4cm}{!}
{\includegraphics{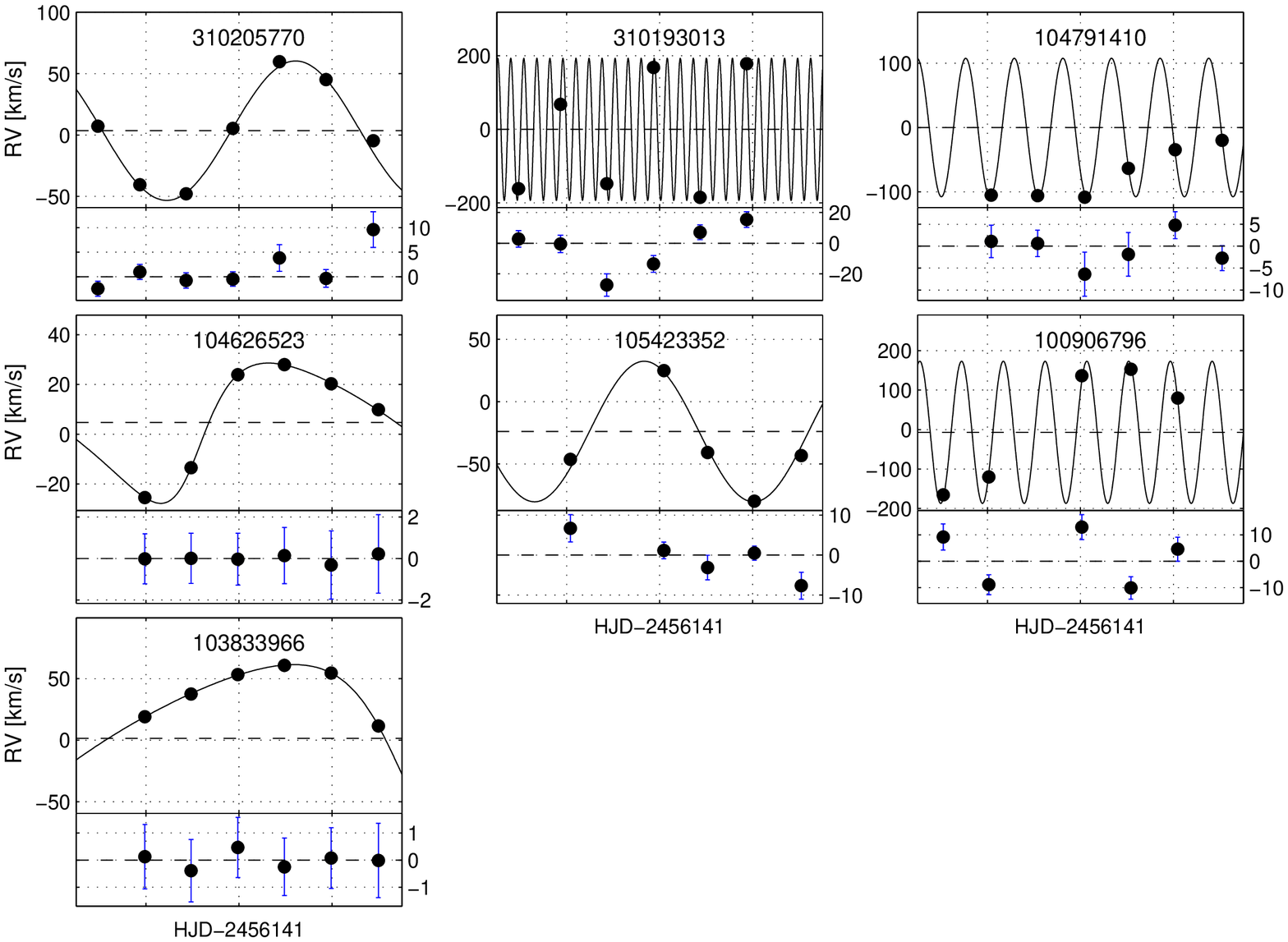}}
\caption{AAOmega RVs (black circles) and the best-fit Keplerian models (solid lines) of the variable components in the confirmed BEER diluted binaries listed in Table \ref{t3h}.}
\label{Fig8}
\end{figure*}

\begin{figure*}[!htp]
\resizebox{18.4cm}{!}
{\includegraphics{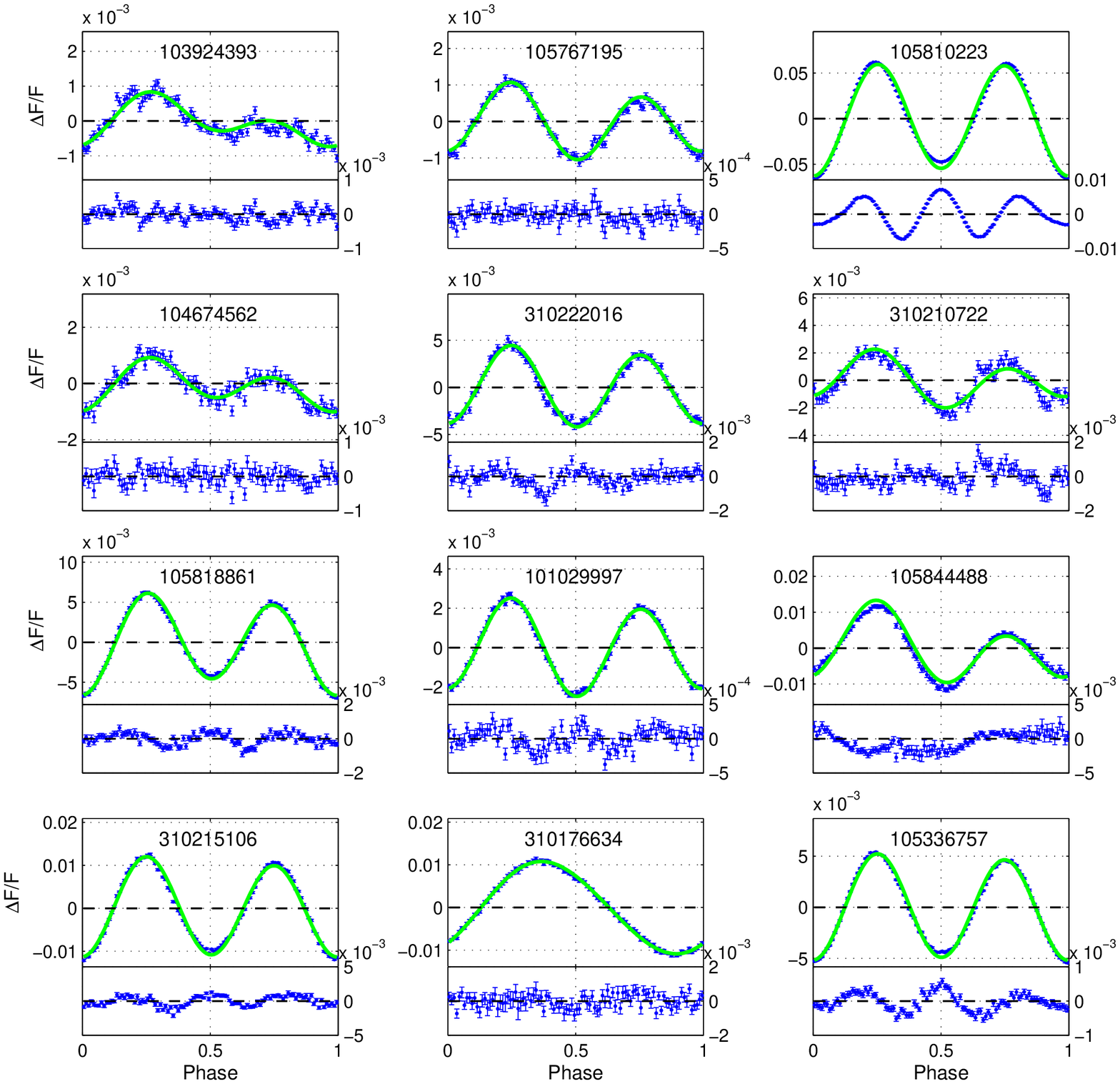}}
\caption{Phase-folded and binned light curves (blue) and the best-fit BEER models assuming a circular orbit (green) of the $70$ BEER binaries confirmed by AAOmega.}
\label{Fig0}
\end{figure*}

\addtocounter{figure}{-1}

\begin{figure*}[!htp]
\resizebox{18.4cm}{!}
{\includegraphics{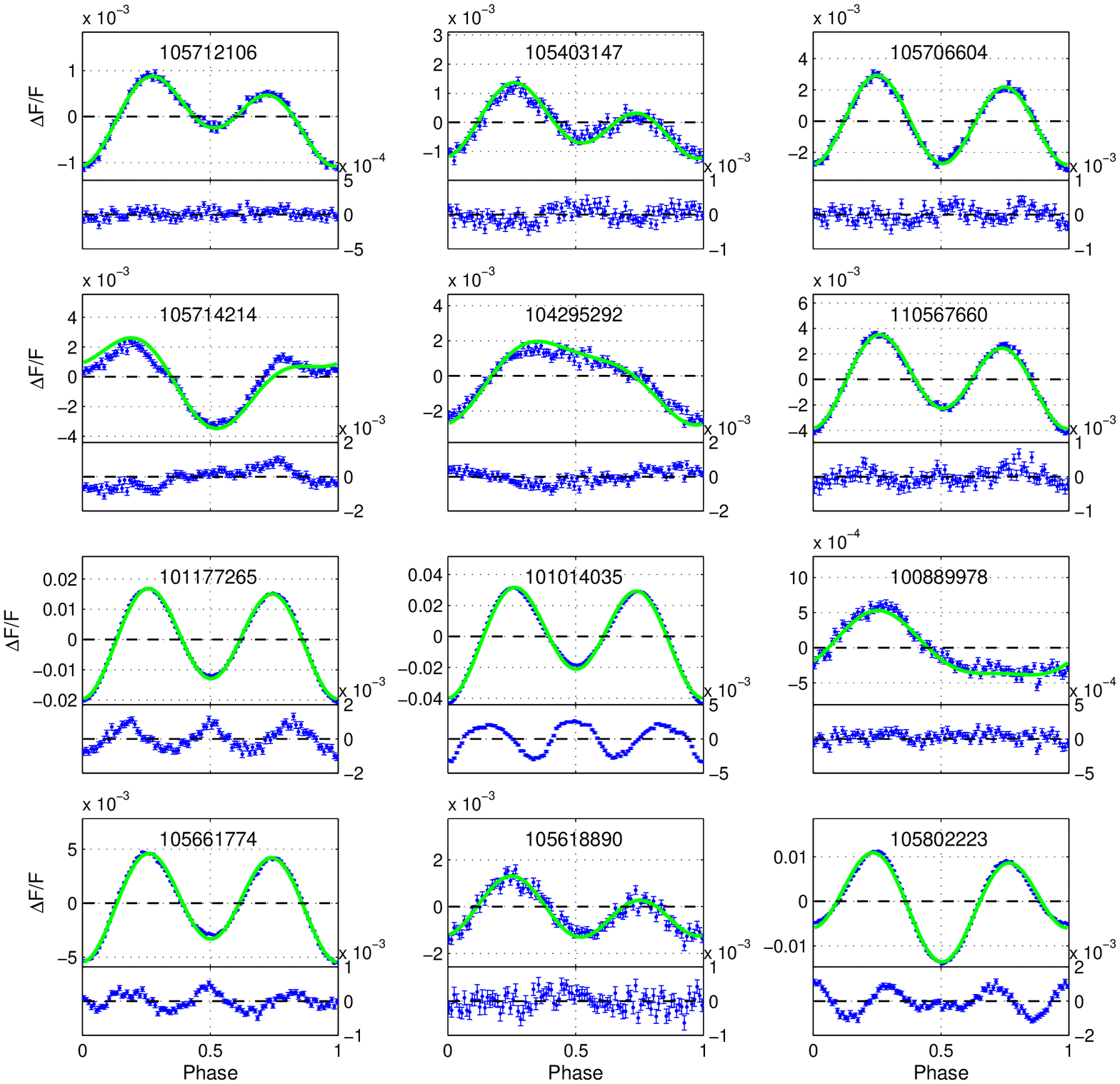}}
\caption{Continued.}
\label{Fig0}
\end{figure*}

\addtocounter{figure}{-1}

\begin{figure*}[!htp]
\resizebox{18.4cm}{!}
{\includegraphics{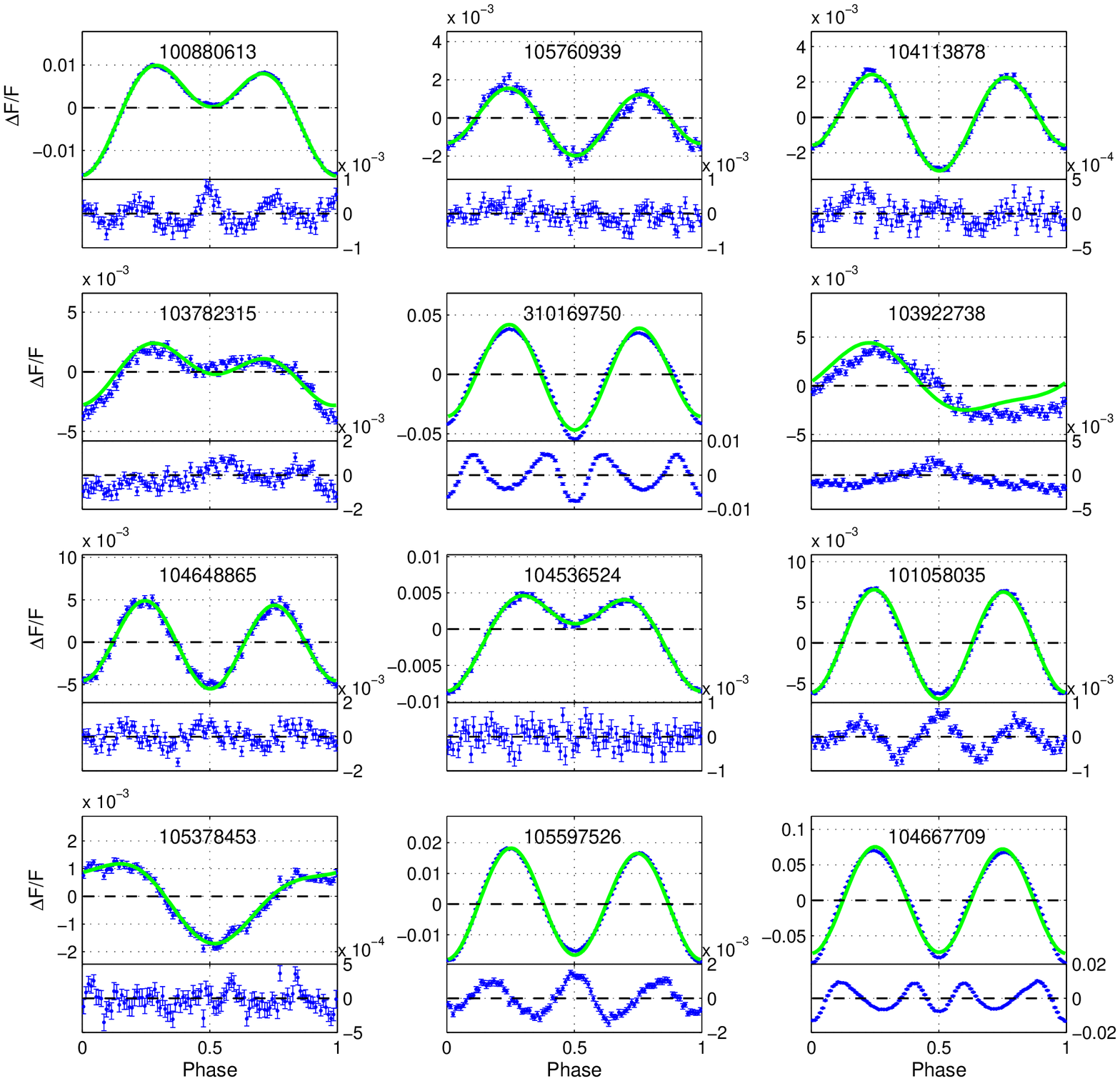}}
\caption{Continued.}
\label{Fig0}
\end{figure*}

\addtocounter{figure}{-1}

\begin{figure*}[!htp]
\resizebox{18.4cm}{!}
{\includegraphics{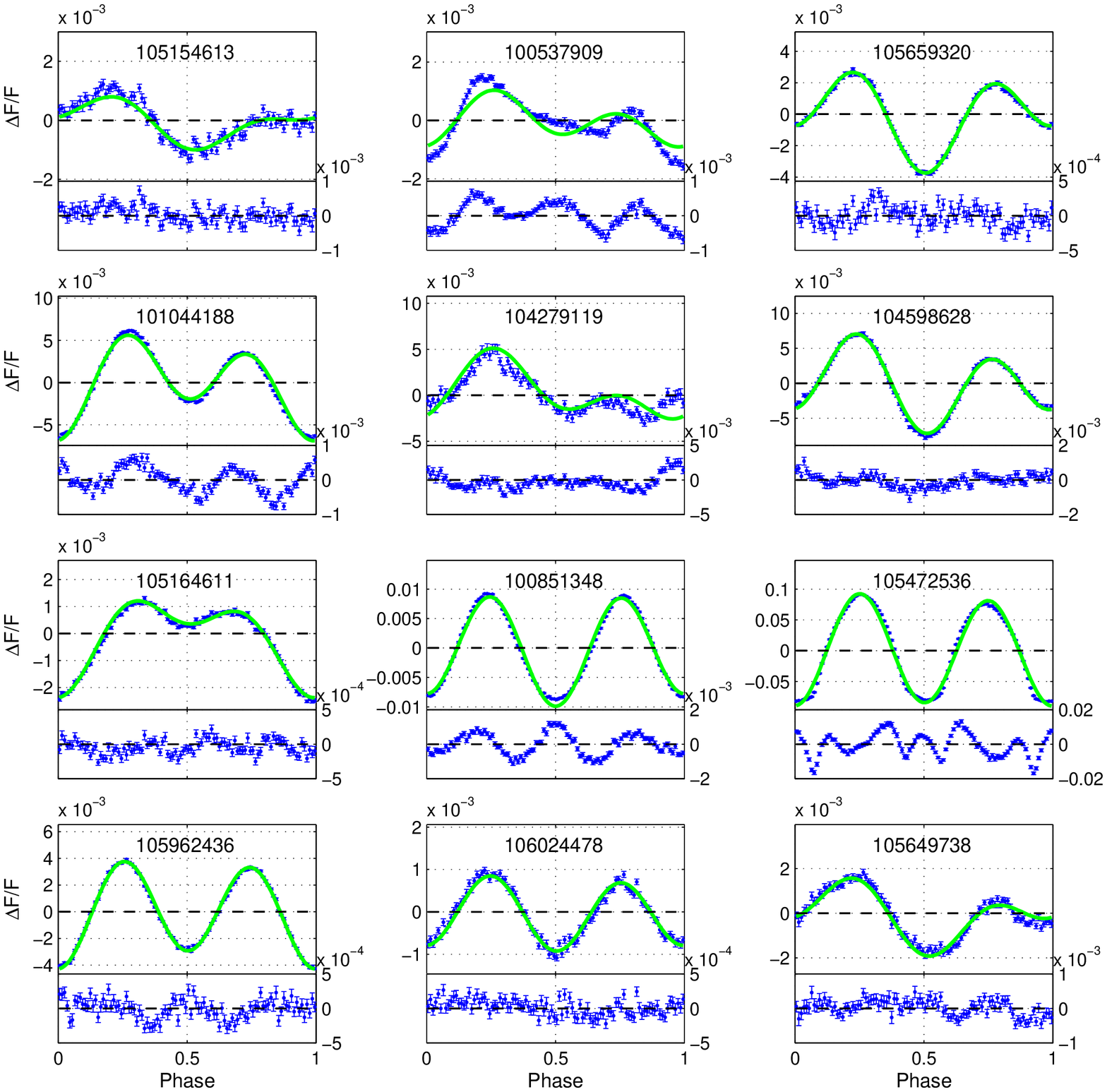}}
\caption{Continued.}
\label{Fig0}
\end{figure*}

\addtocounter{figure}{-1}

\begin{figure*}[!htp]
\resizebox{18.4cm}{!}
{\includegraphics{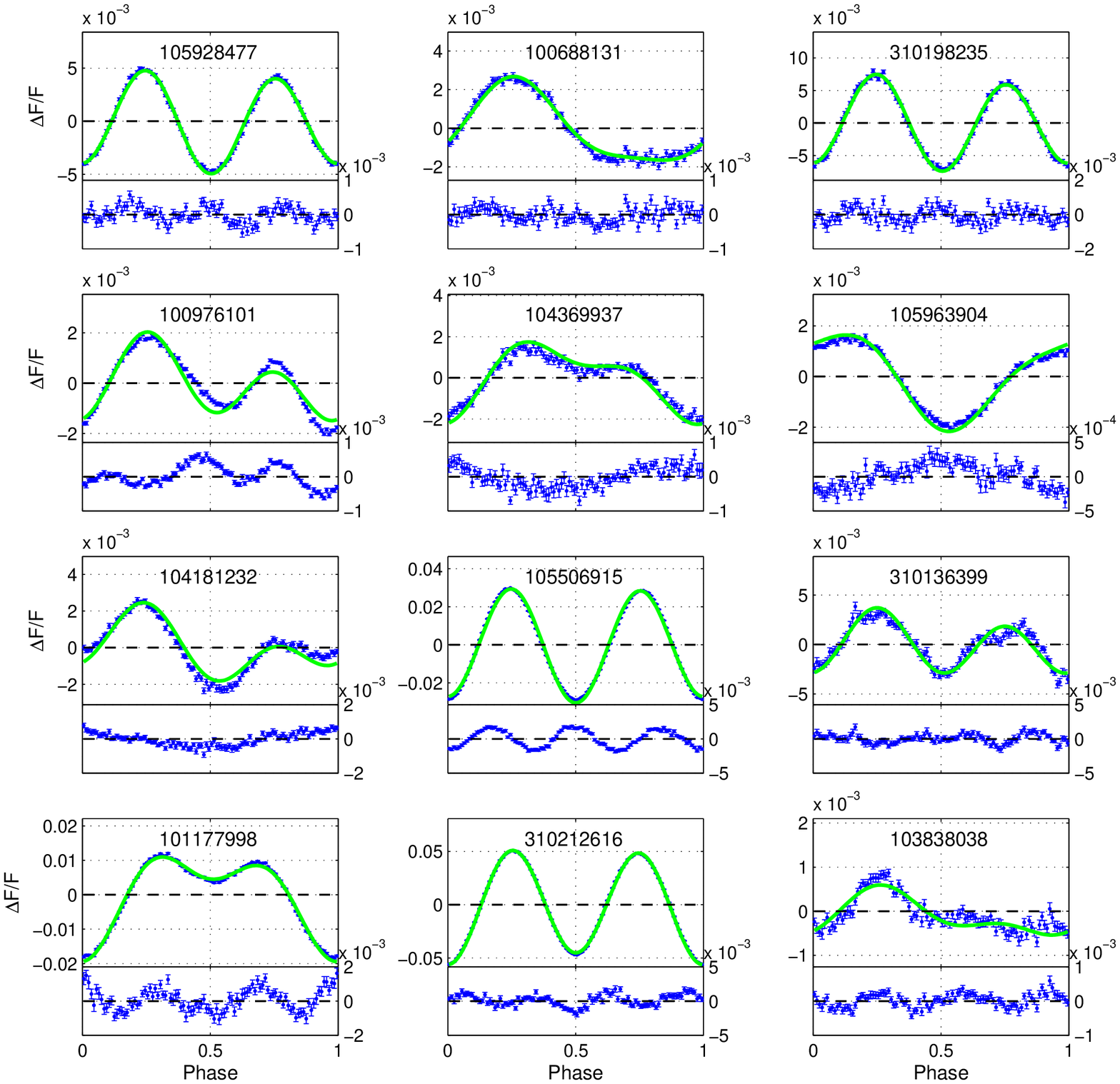}}
\caption{Continued.}
\label{Fig0}
\end{figure*}

\addtocounter{figure}{-1}

\begin{figure*}[!htp]
\resizebox{18.4cm}{!}
{\includegraphics{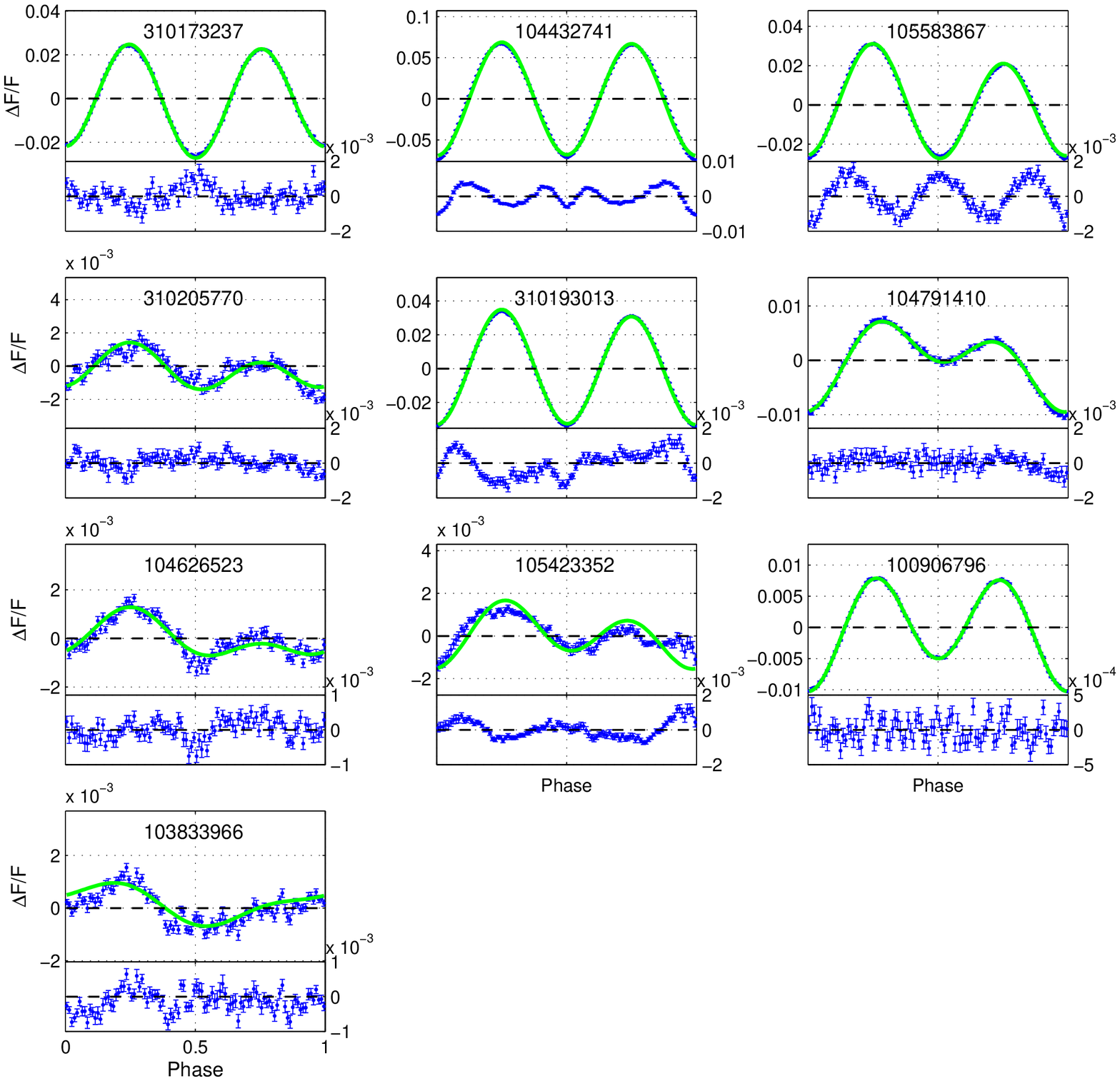}}
\caption{Continued.}
\label{Fig0}
\end{figure*}


\end{document}